\pdfoutput=1
\documentclass{article}

\PassOptionsToPackage{numbers}{natbib}

\usepackage[final]{neurips_2021}

\usepackage[utf8]{inputenc} 
\usepackage[T1]{fontenc}    
\usepackage{hyperref}       
\usepackage{url}            
\usepackage{booktabs}       
\usepackage{amsfonts}       
\usepackage{nicefrac}       
\usepackage{microtype}      
\usepackage{xcolor}         

\usepackage{graphicx}
\usepackage{subcaption}
\usepackage[ruled,vlined]{algorithm2e}
\SetKwInput{KwInit}{Init}
\usepackage[acronym,nomain,nonumberlist]{glossaries}
\makeglossaries
\newacronym{marl}{MARL}{Multi-Agent Reinforcement Learning}
\newacronym{pgg}{PGG}{Public Goods Games}
\newacronym{rl}{RL}{Reinforcement Learning}
\newacronym{crd}{CRD}{Collective Risk Dilemma}
\newacronym{ipd}{IPD}{Iterated Prisoners' Dilemma}
\newacronym{egt}{EGT}{Evolutionary Game Theory}
\newacronym{sos}{SOS}{Stable Opponent Shaping}
\newacronym{lola}{LOLA}{Learning with Opponent Learning Awareness}

\title{Learning Collective Action under Risk Diversity}

\author{%
  Ramona Merhej \\
  ISIR, CNRS, Sorbonne University,  France\\
  Instituto Superior Tecnico, Portugal \\
  \texttt{ramona.merhej@tecnico.ulisboa.pt} \\
  \And
  Fernando P. Santos \\
  Informatics Institute \\
  University of Amsterdam, The Netherlands \\
  \texttt{f.p.santos@uva.nl} \\
  \And
  Francisco S. Melo \\
  INESC-ID and Instituto Superior Tecnico\\
  Universidade de Lisboa, Portugal \\
  \texttt{fmelo@inesc-id.pt}
  \And
  Mohamed Chetouani \\
  ISIR, CNRS UMR 7222, \\
  Sorbonne University, France \\
  \texttt{mohamed.chetouani@upmc.fr}
  \And
   Francisco C. Santos \\
  INESC-ID and Instituto Superior Tecnico\\
  Universidade de Lisboa, Portugal \\
  \texttt{franciscocsantos@tecnico.ulisboa.pt} \\
  \\
}

\begin{document}

\maketitle

\vspace{-1pt}
\begin{abstract}
Collective risk dilemmas (\acrshort{crd}s) are a class of $n$-player games that represent  societal challenges where groups need to coordinate to avoid the risk of a disastrous outcome.
Multi-agent systems incurring such dilemmas face difficulties achieving cooperation and often converge to sub-optimal, risk-dominant solutions where everyone defects.
In this paper we investigate the consequences of risk diversity in groups of agents learning to play \acrshort{crd}s. We find that risk diversity places new challenges to cooperation that are not observed in homogeneous groups. 
We show that increasing risk diversity significantly reduces overall cooperation and hinders collective target achievement.
It leads to asymmetrical changes in agents' policies --- i.e. the increase in contributions from individuals at high risk is unable to compensate for the decrease in contributions from individuals at low risk --- which overall reduces the total contributions in a population.
When comparing \acrshort{rl} behaviors to rational individualistic and social behaviors, we find that \acrshort{rl} populations converge to fairer contributions among agents.
Our results highlight the need for aligning risk perceptions among agents or develop new learning techniques that explicitly account for risk diversity.
\end{abstract}



\section{Introduction}
\label{sec:introduction}

The World Economic Forum recently (January 2021) published its $16^{th}$ report on global risks \cite{websitewef}.
Among the most concerning risks are climate change, biodiversity loss, extreme weather, as well as societal division and economic fragility. 
While it is evident that large collective efforts are needed to avoid these disasters, people, institutions or countries remain reluctant to cooperate. On the one hand, no entity alone has the power of saving the system on its own. 
This is known as the the problem of many hands (PMH) and 
is amplified when actions are not directly harmful but only create the risk of a harm \cite{van2013risk}. On the other hand, cooperation in such contexts entails a social dilemma: the best individual outcome occurs when others contribute to the collective good and risks are avoided without one's intervention. This selfish reasoning, and the shifting of responsibility onto others, configures the so-called Tragedy of the Commons. The tension within individuals/entities created by the urgent need of cooperation, the  individually rational choice to defect, and the uncertainty about future outcomes, makes decision making non-trivial \cite{axelrod1980effective, axelrod1981evolution, kollock1998social}. The \acrfull{crd} is a simple game metaphor that tries to capture such challenges \cite{domingos2020timing, milinski2008collective,santos2011risk,santos2019outcome,vasconcelos2013bottom,vasconcelos2014climate}. 

In a \acrshort{crd}, agents decide how much of their wealth to contribute to a common cause in order to avoid the risk of a future disaster.
The future disaster is only avoided with certainty if the agents manage to collect more contributions than a given target threshold.
The behaviors of individuals playing \acrshort{crd}s have been analyzed both experimentally \cite{dannenberg2011coordination, domingos2020timing, milinski2008collective, tavoni2011inequality} and theoretically, resorting to evolutionary game theory \cite{santos2019outcome, santos2021dynamics, santos2011risk, santos2012evolutionary, vasconcelos2014climate} and multi-agent reinforcement learning \cite{domingos2021pbrl, merhej2021cooperation}. Previous works however, assume an identical risk factor for all agents \cite{domingos2020timing, milinski2008collective, santos2021dynamics, santos2011risk}. In reality, heterogeneous perceptions and exposures to risk are ubiquitous. Most recently, the COVID-19 crisis highlighted our strength and weaknesses in successfully cooperating under such discrepancies. 
Particularly, it showed how different countries adopted different safety measures depending on how risky they assessed the situation to be \cite{alanezi2021comparative, gibney2020whose}.
Diversity in risk perception was not only observed on a national scale but also within each country \cite{lamarche2020socially}. 
The pandemic also revealed how age or medical conditions can result in different levels of risk exposure to a same virus \cite{websitewho}.
Still, some studies have looked into other types of heterogeneities among agents and have reported significant changes in reached cooperation and target achievement \cite{hauser2019social, merhej2021cooperation, vasconcelos2014climate}.
The findings on other heterogeneities motivated us to investigate the effect of introducing risk diversity in a population of agents facing collective risks. 
We examine how averaging out the risk value instead of considering risk diversity can alter the results we observe.

While the game tensions play a decisive role in the choices made by the agents, the final equilibrium of the system also depends on the decision making process of agents.
Decision making can be either modeled as a static or a dynamic process.
A static perspective often models agents as rational and having full knowledge of all possible strategy profiles and outcomes for players. 
This leads agents to converge to the intersection of their best responses, known as the Nash equilibrium.
Yet, experimental studies have shown that humans often make far from rational choices \cite{erev2014maximization, macy2002learning, fudenberg1998theory, skyrms2010signals}, and seem to adapt their policies based on previous experience.
\acrfull{rl} suggests new tools to model decision making dynamics and, in fact, was shown to accurately model human behaviors in social dilemmas \cite{roth1995learning}.
\acrshort{rl} has rapidly evolved in the past years, and several variations were developed specifically to promote cooperation in social dilemmas \cite{foerster2018learning, jacq2019foolproof}. 
We do not use any of these algorithms in our work as they require a large sharing of information and have therefore mostly been applied to $2$-player games.
Additionally, our goal in this paper is to first understand how simple reinforcement learning dynamics can influence agents' cooperation in the presence of risk diversity.
Examining the cooperation challenges that \acrshort{rl} dynamics may pose under risk diversity is essential before moving on to designing algorithms that solve these challenges.
As such, we focus on independent reinforcement learning algorithms where agents can only observe their own actions and rewards.

To assess the strength and weakness of adaptive agents in reaching cooperative solutions under risk diversity, we compare the learned strategies with a set of static solutions. 
On one hand, we compare the behaviors under \acrshort{rl} to those prescribed by individualistic and rational game theory, and on the other hand, to socially optimal solutions that maximize the total welfare in the population.

We begin our paper with Section~\ref{sec:related work} on related work. 
After that, in Section~\ref{sec:model} we model the collective risk dilemma, explain how risk diversity is introduced, and describe agents' learning dynamics.
This is followed by Section~\ref{sec:risk ineq:static study} in which we derive the static solutions for the game.
We display our results in Section~\ref{sec:results:risk diversity} and conclude our work in Section~\ref{sec:risk ineq:conclusion}.


\section{Related work}
\label{sec:related work}

We examine how in a population of adaptive agents facing collective dilemmas, risk diversity can affect that population's ability to cooperate and effectively avoid a disastrous outcome. 
Previous works on \acrshort{crd}s, both experimental and theoretical, have concluded that higher risk translates into higher cooperation and consequently may help in escaping the tragedy of the commons \cite{merhej2021cooperation, milinski2008collective, santos2011risk, santos2012evolutionary}.
But the global risk is not the only decisive factor in an agent's willingness to cooperate.
The introduction of different inequalities between agents can have a significant impact on cooperation.
Under evolutionary game theory, inequalities in wealth, productivity and benefits are found to reduce agents' cooperation in a continuous public goods game \cite{hauser2019social}.
Similar results are also found in a threshold public goods game when agents can only adapt by imitating agents from the same wealth class \cite{vasconcelos2014climate}. 
Wealth inequality is also shown to hinder target achievement in a study on \acrshort{crd}s with reinforcement learners \cite{merhej2021cooperation}, and in an experimental study on a threshold public goods games \cite{tavoni2011inequality}.

While most studied heterogeneities in the literature focus on wealth inequality, we argue that risk diversity is another heterogeneity worth studying in populations facing collective risks.
We distinguish between two types of risk diversity: risk perception diversity -- where agents perceive a same risk as higher or lower than it actually is  -- and risk exposure diversity -- where some agents are more or less vulnerable to facing a risk.
In the context of risk perception diversity, a survey of 119 countries confirms significant variance in public concern and risk assessment of the global climate change problem \cite{lee2015predictors}.
In the context of risk exposure diversity, we saw that the recent COVID-19 pandemic led to the distinction between people at normal risk and those at increased risk for severe illness from COVID-19 \cite{websitewho}.
Governmental units such as the Occupational Safety and Health Administration (OSHA) of the United States Department of Labor have classified jobs into four potential risk exposure levels 
\cite{websiteosha}.
The Organization for Economic Co-operation and Development (OECD) published a document urging governments to support the most vulnerable people \cite{websiteoecd}
and several other studies on that subject have been published in different countries \cite{benfer2020health, osama2020protecting, van2020covid}.

Risk diversity is therefore a fundamental feature taken into account by countries when elaborating their safety measures and preventive policies.
Risk diversity on an individual level, also translates into behavioral diversity and safety measure compliance \cite{thanh2020survey}.
Whether risk diversity emerges from exposure or perception diversity is irrelevant when studying the emergent agent policies and hence their ability to reach a target threshold.
However, it is the resulting consequences of reaching or not the threshold that differ if agents are effectively at higher/lower risk, or if this discrepancy is merely an impression.
The behavioral differences between people at high risk and those not, as well as the previous work that demonstrated the impact of heterogeneities on a population's cooperation capabilities, inspired us to dedicate a study on risk diversity in collective risk dilemmas.

Although $n$-player non symmetrical social dilemmas and games with mixed-motives are abundant in the real world, cooperation in multi-agent reinforcement learning has mainly focused on $2$-player games.
A study on sequential social dilemmas with deep \acrshort{rl} \cite{leibo2017multi}, identified coordination sub-problems that prevent proper cooperation of agents. Coordination problems in \acrshort{marl} are quite common and are not only restricted to social dilemmas \cite{matignon2012independent}.
One of the reasons for coordination difficulty in \acrshort{marl} is the non-stationarity of the opponent and the simultaneous policy updates of the players \cite{balduzzi2018mechanics}. Suggested solutions try to increase agents' \textit{understanding} of the opponent's dynamics and leverage these to achieve higher cooperation. 
One algorithm proposes predicting the opponent's policy changes before computing the agent's policy gradient \cite{zhang2010multi}. 
Another alternative suggests differentiating through the variations of the opponent to actively shape their learning 
\cite{foerster2018learning}. 
A third solution incorporates both policy prediction and opponent shaping to increase stability while simultaneously escaping saddle points \cite{letcher2018stable}. 


Other solutions to increase cooperation in \acrshort{marl} focus on enabling \textit{communication} capacities between agents.
Communication can take several forms. For example, agents may communicate by sending messages \cite{foerster2016learning}, sharing intentions \cite{kim2021communication} or experiences \cite{christianos2020shared}, advising actions \cite{omidshafiei2019learning} to one another etc. Enriching agents with communication capabilities has shown to improve performance \cite{foerster2016learning, christianos2020shared}, speed up learning \cite{omidshafiei2019learning, foerster2016learning, christianos2020shared} and enhance coordination \cite{kim2021communication, omidshafiei2019learning}. Implementing a centralized critic with decentralized actors is another form of indirect communication and information sharing among agents that can increase performance and cooperation \cite{baker2019emergent, foerster2018counterfactual, lowe2017multi}.

A third set of solutions for overcoming cooperation difficulty in \acrshort{rl} introduce \textit{conditional commitment} in agents' policies. One example is an algorithm designed to always asymptotically behave as a Tit-for-Tat strategy by learning simultaneously a cooperative and a selfish \textit{Q}-function and alternating between them to avoid exploitability \cite{jacq2019foolproof}.

Finally, solutions modifying agents' motivations can be seen as \textit{institutional} solutions \cite{dafoe2020open}. Notably, in \acrshort{marl}, intrinsic rewards can be engineered and added to environmental rewards to help agents solve a sub-problem of the game and facilitate the emergence of coordination \cite{liu2019emergent}. 

We note that all advised solutions for increasing cooperation in \acrshort{marl} settings focus on $2$-player games. Major computational and convergence problems still inhibit the scaling of these algorithms to $n$-player games. Additionally, most solutions are developed to increase cooperation in purely cooperative settings. We propose a non-symmetrical $n$-player mixed-motive game. We describe the emergent behaviors of simple reinforcement learners in these settings. The goal of the paper is to recognize the cooperation challenges of reinforcement dynamics in the context of such dilemmas. The outcomes of our study can be exploited to design more effective \acrshort{rl} algorithms in the future. However, such developments remain out of the scope of our current paper.

\section{Model}
\label{sec:model}


\subsection{Game dynamics}
\label{sec:game definition}

A \acrfull{crd} is a game in which agents need to cooperate to avoid an eventual disaster \cite{domingos2020timing, milinski2008collective,santos2011risk,santos2019outcome,vasconcelos2013bottom,vasconcelos2014climate}.
Agents' success in avoiding the disaster  requires a minimum of collective efforts.
Effort is modeled by the costly contribution of players towards a common pool.
If contributions are below the threshold they will not alleviate the consequences of the disaster.
Additionally, all contributions above the threshold do not create any additional value for the players.
As a result, agents are simultaneously motivated to cooperate to increase the chances of avoiding the disaster, and to defect and free-ride with the hope that others will ensure disaster avoidance.

Formally, in a population of finite size $Z$, we allocate for every player an initial endowment $b$.
 Players are then sampled into groups of size $N$ to play \acrshort{crd}s.
They need to jointly collect enough contributions to reach a target threshold $\mathbf{t}$ to avoid with certainty some common disaster.
If a group manages to achieve the threshold target, the disaster is avoided and players only lose what they had contributed to the common pool.
However, should the target not be met, agents, depending on their level of risk exposure to the disaster $r_i$, will lose a fraction $p$ of their remaining endowment.
At the end of the game, player $i$ who started with an initial endowment $b$ will be left with
\begin{equation}
\begin{aligned}
\label{ind_good}
b_{i}^{final} = \left\{
    \begin{array}{ll}
        (1-c_i)b & \mbox{if the disaster was avoided,} \\
        (1-c_i)b  - p(1-c_i)b & \mbox{if it wasn't.}
    \end{array}
\right.
\end{aligned}
\end{equation}

where 
$c_i$ is a binary choice of contributing $0$ or a fraction $c$ of the endowment to the pool ($c_i \in \{0,c\}$).

The perceived benefits or harm of these losses in endowment is a subjective function known as the \textit{utility} in economic game theory.
One common utility function is the log-utility.
The log-utility function has been used  when studying the impact of wealth inequality in collective risk dilemmas \cite{merhej2021cooperation} and is used more broadly in economy to capture what is known as a diminishing marginal utility  \cite{peters2016evaluating}. 
It supposes that the loss of a given amount of money is perceived as more painful by poorer individuals than by richer ones.
While all agents are equally wealthy in our scenario, we do intend to examine mixtures of heterogeneities in future works, such as the combination of wealth inequality with risk diversity. 
With that in mind, to better compare our results with future works, we decide to also adopt a log-utility function. 
The payoffs of the game are expressed as the difference in the $\log$ of agents' wealth before and after a game was played.
Avoiding a disaster will cost a cooperator $x_C=\log \left( \frac{b-cb}{b} \right) = \log(1-c)$, and a defector $x_D=\log\left(\frac{b}{b}\right)=0$ or nothing.
Facing a disaster will cost cooperators $\bar{x}_C=
\log(1-c-p(1-c))$ and defectors $\bar{x}_D=
\log(1-p) $. 
The necessary conditions on $r$, $c$, $p$ and $\mathbf{t}$ that ensure that the game designed is a social dilemma are detailed in Appendix~\ref{appendix:game definition}.
The goal of each player is to find a probabilistic strategy $\pi_i^*$ - representing the probability of player $i$ choosing to cooperate - that maximizes the payoff.

\paragraph{Introduction of risk diversity}
We consider risk diversity in the form of binary risk \textit{classes}.
That is, we split our population into two classes: agents at high risk of being affected by the disaster and agents at low risk.
The former group represents a fraction $z_H$ of the population, and the latter a fraction $z_L=1-z_H$. 
Given an average population risk value $r$ and a risk diversity value $\delta$, if the target is not achieved, agents at high risk will lose an additional fraction $p$ of their remaining wealth with probability $r_H=r+ \frac{1}{2 z_H} \delta$ while agents at low risk only face that disaster with a risk probability $r_L=r-  \frac{1}{2 z_H} \delta$.


\paragraph{Numerical Values}
\label{par:game param risk diversity}

The population size is set to $Z=200$ individuals.
The agents are organized in groups of $N=6$.
They are given an initial endowment $b=1$ and can choose to either cooperate and contribute a fraction $c=0.1$ of it to a common pool or defect and contribute nothing. 
Participants have stochastic policies $\pi_i$ that define the probabilities of choosing each action.
The threshold $\mathbf{t}$ is set so that the target is only achieved if at least half of the agents in a group cooperate, i.e., $\mathbf{t} = M c b$ with $M=\frac{N}{2}$. 
Agents at high and low risk are equally frequent in the population with $z_H=z_L=50\%$ of the population. 
This means, for an average population risk $r$ and a risk diversity $\delta$, agents at high risk will face a disaster with probability $r_H=r+\delta$ while agents at low risk will face a disaster with a risk probability $r_L=r-\delta$.
If the threshold target is not achieved, every agent that faces a disaster pays a penalty of $p=0.7$ or $70\%$ of its remaining wealth. 
We proceed with two experiments: in the first, we fix the diversity value to $\delta=0.1$ and test varying average risk values $r$, while in the second, we set the population average risk value to $r=0.5$ and vary the risk diversity value $\delta$. This allows us to better understand the impact of risk diversity for regimes of high and low baseline risk ($\delta$ fixed and varying $r$) and also the impacts of increasing symmetric risk diversity ($r=0.5$ and varying $\delta$).


\subsection{Agent learning algorithm}
\label{sec:agent dynamics}

The goal of the paper is to understand how simple reinforcement dynamics can encourage or discourage cooperative behaviors in populations with risk diversity.
We choose to model the agents learning dynamics using the Roth-Erev Algorithm \cite{roth1995learning} which was shown to successfully model human decision making in social dilemmas.
Accordingly, we create a population of $Z$ agents and allow every player $i$, at every timestep $k$, to hold and update a propensity vector that assigns a propensity value for each of the possible actions.
In the $2$-actions collective risk dilemma, this translates to a vector $\mathbf{q}_{i,k} = \begin{bmatrix}q_{i,k}(C), q_{i,k}(D) \end{bmatrix}^T$ where $q_{i,k}(C)$ and  $q_{i,k}(D)$ are the respective propensities for the cooperative and the defective action at timestep $k$.
For every interaction $k$ in the learning process, agents normalize their propensity vector and sample one of the two actions following the obtained probabilities.
At the end of the $k^{th}$ game, when returns are distributed, every player $i$, depending on the selected action $A$ and the received reward $x$, updates the propensity vector such that

\begin{equation}
\label{eq:update rule}
    \begin{aligned}
        &q_{i,k+1}(A) \; \; = (1-\phi)q_{i,k}(A) + x\\
        &q_{i,k+1}(\neg A) = (1-\phi)q_{i,k}(\neg A)
    \end{aligned}
\end{equation}

where $\phi$ is a forgetting  parameter that inhibits the propensities from growing to infinity. Further details about the population training procedure as well as numerical values are given in Appendix~\ref{appendix:agent definition}.


\section{Static study}
\label{sec:risk ineq:static study}

To highlight the peculiarities of adaptive agents in social dilemmas, we compare the learned and adaptive solutions to other statically tailored solutions.
Particularly, we are interested in comparing the learned solutions to 1) solutions that are rational from an individualistic point of view and 2) to solutions that are rational from a communal or collective point of view.
The rational solution from an individualistic perspective is the Nash equilibrium while the rational equilibrium from a communal perspective is the total welfare maximizing solution.

Given the computational difficulty of finding Nash equilibria or social welfare maximizing solutions for an $n$-player, non-linear general-sum game with continuous strategies, we choose to work with class-based solutions that were previously proposed for \acrshort{crd}s under wealth inequality \cite{merhej2021cooperation}.
The solutions pre-impose perfect coordination between players of a same class (here agents at high/low risk), by forcing all agents from a same class to follow the same strategy.
As a result, agents of a same class can be modeled as one large agent which transforms the $n$-player game into a $2$-player game.


\paragraph{Class-based Nash}
\label{subsec:risk ineq:class-based Nash}

Following a similar reasoning to the one detailed for wealth inequality \cite{merhej2021cooperation}, we transform the $n$-player game into a $2$-player game.
After transformation of the game, the method relies on a graphical approach to extract the intersection points of the best response strategies of the two classes. 
The intersection points represent the class-based Nash equilibria.
We repeat this to extract class-based Nash solutions for all game settings and all risk diversities.
The exact changes in reasoning with respect to wealth inequality as well as the plots for extracting class-based Nash points can be found in Appendix~\ref{appendix:subsec:risk ineq:class-based Nash}.
We use results obtained under class-based Nash equilibrium as the baseline to evaluate how rational the learned strategies of adaptive reinforcement learners are. 


\paragraph{Class-based maximum welfare}
\label{subseb:risk ineq:max welfare}

We extend the class-based Nash method, to extract class-based maximum welfare points.
We continue to impose absolute equality and fairness within a given class and evaluate the total secured welfare of a population for different combinations of strategies. 
Here, instead of plotting best response lines, we draw a heat-map with the secured welfare for each combination of class strategies.
We define as the class-based maximum welfare solution, the point that minimizes the total losses in welfare for the population.
Further details and the corresponding heat-maps are given in Appendix~\ref{appendix:subseb:risk ineq:max welfare}.


\section{Results}
\label{sec:results:risk diversity}

We study the consequences of risk diversity in populations of \acrshort{rl} agents learning to play \acrshort{crd}s.
After the training phase, the strategies are evaluated based on the resulting population's probability of achieving the target threshold $\mathbf{t}$.
For every setting, we rollout a game where the population is split into groups of $N$ players.
In each group, agents, following their learned strategies, choose to either contribute or not.
We define as $\eta$, the average percentage of groups in the population that reach the target threshold. 
This random variable is evaluated and averaged over $10^6$ simulations.
Studies are run both on heterogeneous populations with risk diversity, as well as on their homogeneous counterparts (i.e., populations with the same average risk factor $r$ but no diversity $\delta$). 


\subsection{Effect of risk inequality on cooperation levels and target achievement}
\label{subsec:results:effect of risk inequality}

\begin{figure*}[h!]
 \centering
 \begin{subfigure}[t]{0.32\textwidth}
   \includegraphics[width=\textwidth]{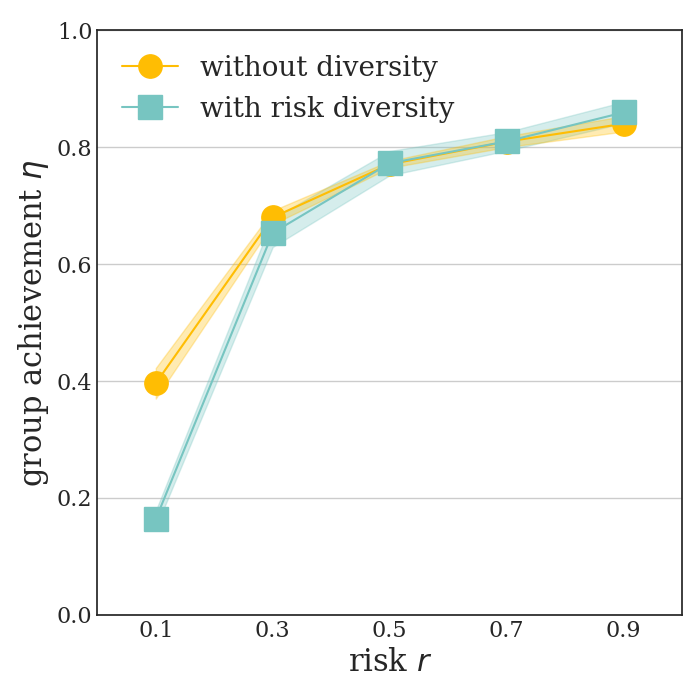}
 \caption{}
 \label{fig:with without ineq risk:eta}
 \end{subfigure}
 \begin{subfigure}[t]{0.32\textwidth}
   \includegraphics[width=\textwidth]{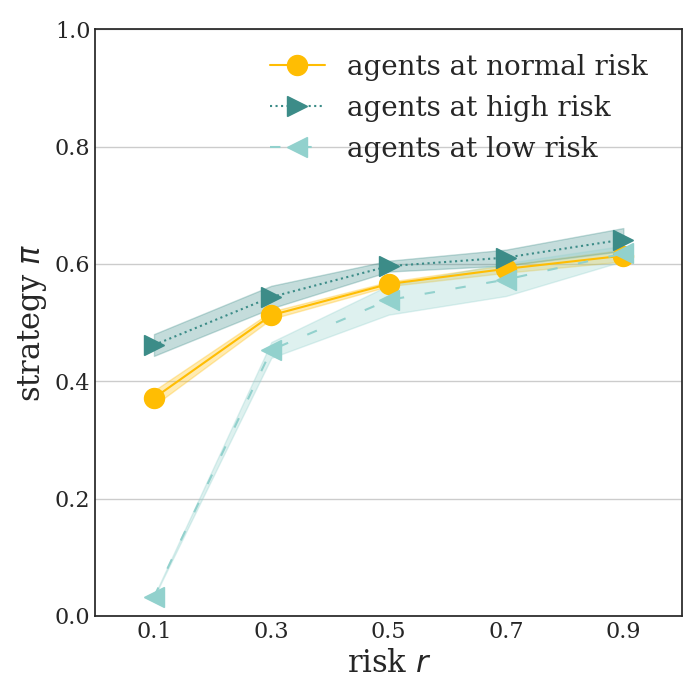}
 \caption{}
 \label{fig:with without ineq risk:pi}
 \end{subfigure}
 \caption{\textbf{\acrshort{rl} dynamics:} (\subref{fig:with without ineq risk:eta}) Group achievement of populations with and without risk diversity w.r.t. the risk factor $r$. (\subref{fig:with without ineq risk:pi}) Learned strategies of agents from a population without diversity and those of agents at high and low risk from a population with diversity w.r.t. the game's risk $r$. 
 Shaded areas represent the standard deviation over 5 runs.}
\label{fig:with without ineq risk}
\end{figure*}

To study the effect that risk inequality can have on a population facing a collective risk dilemma, we begin by comparing the group achievement rate $\eta$ and the learned strategies of a homogeneous population on one hand, with those of a heterogeneous population with risk diversity factor $\delta=0.1$ on the other hand.
We plot the results for varying average risk factors $r$ in Figure~\ref{fig:with without ineq risk}.
With or without inequalities, we observe that the group achievement rate increases with the risk factor $r$ (Figure~\ref{fig:with without ineq risk:eta}) as a result of higher cooperation willingness (Figure~\ref{fig:with without ineq risk:pi}) when the costs of failure increase. 
These results are consistent with other studies on collective risks \cite{merhej2021cooperation, santos2011risk}.

However, while most studies reported that inequalities had a decisive impact on group achievement \cite{merhej2021cooperation, santos2008social, vasconcelos2014climate}, risk diversity has little or no impact on group performance for all risk values of $r \geq 0.3$. Additionally, for $r \geq 0.3$, only minor differences are observed in the strategies of agents at high and low risk, which as $r$ increases, converge to the strategies learned by a homogeneous population. 
The results are in contrast with the results for $r=0.1$ where risk diversity reduces target achievement and causes and a large gap in cooperation between the two classes.
As $r$ increases, the relative strength of the diversity $\delta/r$ decreases resulting in more homogeneous behaviors between the classes.

\begin{figure*}[h!]
 \centering
 \begin{subfigure}[t]{0.32\textwidth}
   \includegraphics[width=\textwidth]{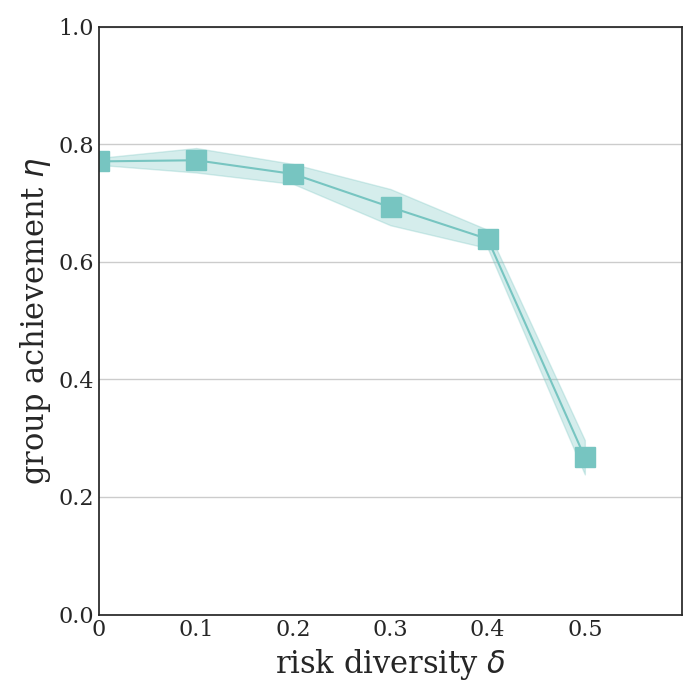}
 \caption{}
 \label{fig:ineq risk with delta:eta}
 \end{subfigure}
 \begin{subfigure}[t]{0.32\textwidth}
   \includegraphics[width=\textwidth]{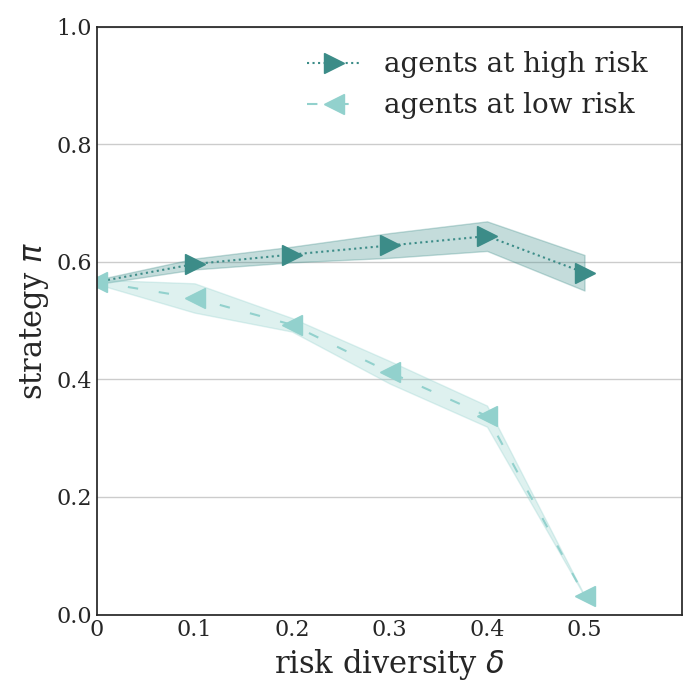}
 \caption{}
 \label{fig:ineq risk with delta:pi}
 \end{subfigure}
 \caption{\textbf{\acrshort{rl} dynamics:} (\subref{fig:ineq risk with delta:eta}) Overall group achievement of populations with varying risk diversity factors $\delta$. (\subref{fig:ineq risk with delta:pi}) Learned strategies of agents at high and low risk in populations of different risk diversity factors $\delta$. 
Shaded areas represent the standard deviation over 5 runs. $\delta=0$ represents populations without risk diversity. The average risk in all populations is $r=0.5$.}
\label{fig:ineq risk with delta}
\end{figure*}

Upon that, we investigate the role of the diversity factor $\delta$.
In a second experiment, we fix the average risk in the population to $r=0.5$ and evaluate populations of varying risk diversity factors $\delta$. 
In Figure~\ref{fig:ineq risk with delta:eta}, we observe how for the same average risk, stronger diversity causes a drop in achievement. The steepest drop occurs when all the population's risk is only carried by half of the population, i.e. for $\delta=0.5$ ($r_L=r-\delta=0$).
Figure~\ref{fig:ineq risk with delta:pi} shows the strategies followed by individuals at high and at low risk in each of the populations. We notice an increased gap in cooperation between the two classes as one class adjusts its cooperation rate faster than the other one. 
The reduced cooperation of agents at low risk is not compensated by a similar increase in cooperation from agents at high risk which explains the drop in target achievement as the $\delta$ increases.

\subsection{Adaptive vs. class-based static strategies}
\label{subsec:results:adaptive vs static}

\begin{figure*}[h!]
 \centering
\begin{subfigure}[t]{0.33\textwidth}
   \includegraphics[width=\textwidth]{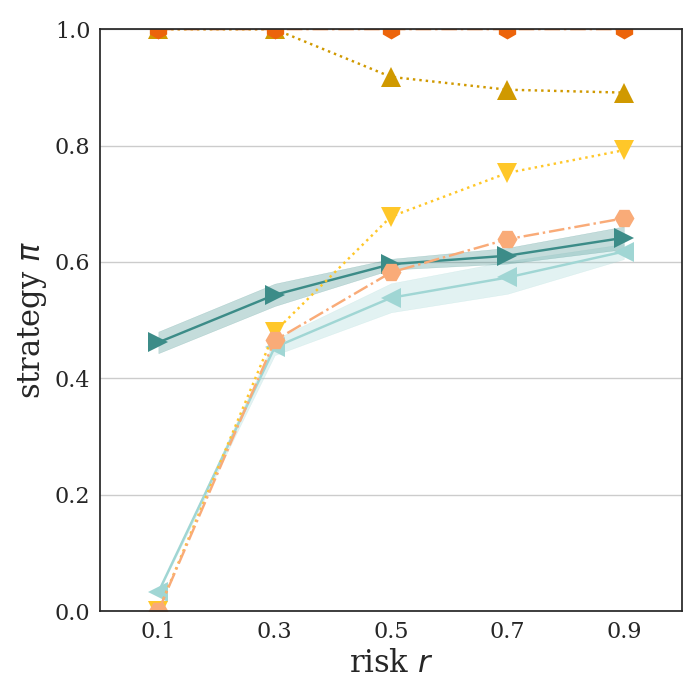}
  \caption{}
  \label{fig:adaptive vs static risk:risk:pi}
  \end{subfigure}
 \begin{subfigure}[t]{0.33\textwidth}
   \includegraphics[width=\textwidth]{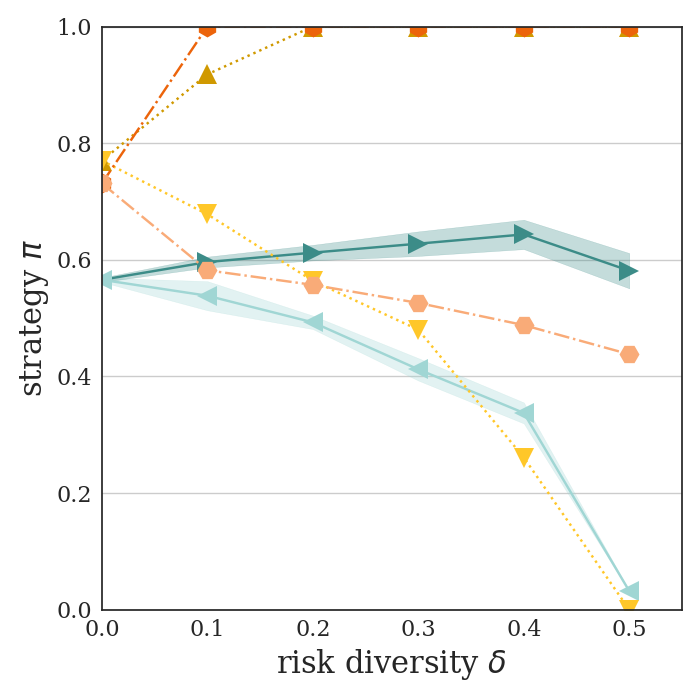}
 \caption{}
 \label{fig:adaptive vs static risk:delta:pi}
 \end{subfigure}
  \begin{subfigure}[t]{0.33\textwidth}
   \includegraphics[width=\textwidth]{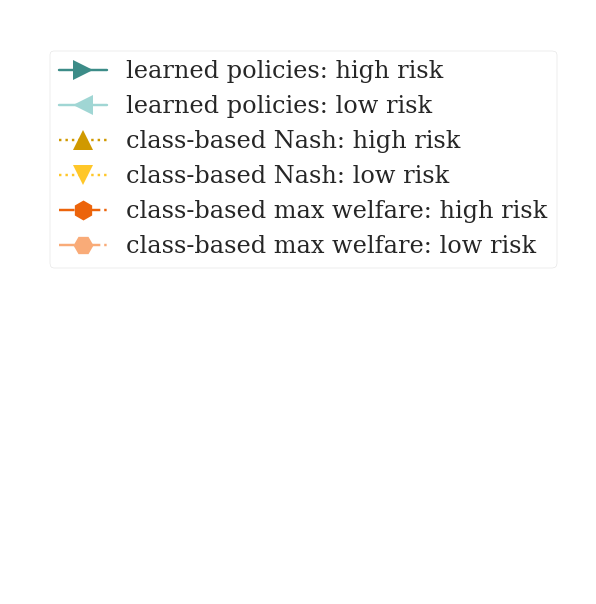}
 \caption*{}
 \label{}
 \end{subfigure}
 \caption{\textbf{Adaptive vs static dynamics:}  Strategy profiles of agents at high and low risk in an adaptive \acrshort{rl} population compared to statically extracted strategies, (\subref{fig:adaptive vs static risk:risk:pi}) for $\delta=0.1$ and different risk factors $r$, and
 (\subref{fig:adaptive vs static risk:delta:pi}) for $r=0.5$ and different risk diversity factors $\delta$.}
 \label{fig:adaptive vs static risk}
 \end{figure*}

Next, we explore the effect of risk diversity on the policies of \acrshort{rl} agents compared with a baseline of 1) individualistically rational solutions and 2) socially rational solutions.
These are respectively the class-based Nash and the class-based maximum welfare policies from Section~\ref{sec:risk ineq:static study}. 
First, in Figure~\ref{fig:adaptive vs static risk:risk:pi}, we consider a constant diversity $\delta=0.1$ and observe the effect of changing the risk.
Then, in Figure~\ref{fig:adaptive vs static risk:delta:pi}, we fix the average risk $r=0.5$ and look at the impact of varying diversity.
In all cases, we notice that adaptive agents converge to more egalitarian solutions than the class-based agents in the sense that the gap in cooperation between agents at high and low risk is consistently smaller for \acrshort{rl} populations compared to class-based populations.
A higher risk reduces this gap while a higher diversity increases it.

When comparing class-based Nash to class-based social welfare maximizing solutions, we notice that for agents at low risk, selfish Nash solutions usually recommend higher cooperation than social welfare solutions. 
However, in Figure~\ref{fig:adaptive vs static risk:delta:pi}, we observe a cross point between the class-based Nash solutions and the class-based social welfare maximizing ones at $\delta=0.2$. From a selfish perspective, as $\delta$ increases, agents at low risk become less exposed to the disaster and the costs of high cooperation become larger than the costs of failure.
In contrast, from a social perspective, as $\delta$ increases, the losses on agents at high risk increase and agents at low risk need to pitch-in to avoid further losses on the population.
\acrshort{rl} agents at low risk learn behaviors similar to the class-based Nash solutions and eventually stop cooperating with increasing diversity. 
However \acrshort{rl} agents at high risk have trouble converging to solutions of high cooperation as recommended by the class-based Nash policies.
  
\subsection{Nash vs. class-based Nash}

On a final note, we highlight the distinctions between the class-based Nash and the general Nash solution.
The general Nash is a point where no agent can increase its payoff by deviating \textit{alone} from the chosen strategy. In other words, the Nash equilibrium considers fully independent players with no pre-established coordination. It finds a solution for both the game's cooperation and coordination dilemmas.
The class-based Nash however, supposes no agent ever deviates alone from a chosen strategy. Instead, all agents of a same class, move in the same coordinated manner. The class-based Nash reduces the degrees of freedom and only solves the game's inter-class cooperation dilemma.

As a result, the class-based Nash and the Nash equilibria may not always converge to the same solution.
For instance, total defection is a Nash equilibrium in the \acrshort{crd}:
if $Z-1$ agents in the population defect, then the $Z^{th}$ agent's best response is to defect as well since the target threshold cannot be achieved alone.
Yet, this equilibrium point was not found in any class-based Nash solutions. 
As a result of moving collectively, defection is less desirable for an agent because it simultaneously implies a defection of the rest of the agents in the same class. In all class-based Nash strategies, if we fix $Z-1$ strategies in the population and only allow one agent to change its strategy, defection is indeed the most profitable choice. 
This proves that the class-based Nash points are not Nash equilibria.

Interestingly, if we study the learned strategies for all tested risk values, i.e., if we fix $Z-1$ strategies in the population and only allow the $Z^{th}$ agent to change its strategy, defection is again the most profitable choice. Learned strategies are not Nash equilibria either.
We hypothesize that the large size of the population can hamper convergence to Nash equilibria for adaptive agents. 
In a similar way to the class-based update, if several agents in a population simultaneously increase their defection rate, the next interaction may become less profitable as the increase in failure (caused by a reduction in target achievement) is not compensated by the individual decrease in cooperation cost. 
Assessing the benefits of diverging alone from a strategy profile, which is necessary for computing Nash-equilibria, is not easily done in \acrshort{rl} populations  where all agents can simultaneously change their strategies. 
Learning with \acrshort{rl} in large populations seems to help in escaping defective Nash equilibria.


\section{Conclusion and discussion on cooperative capabilities}
\label{sec:risk ineq:conclusion}

We examined how risk inequality between \acrshort{rl} agents can affect a population's target achievement rate and the cooperation levels of different risk classes.
First, we found that high risk diversity causes a noticeable decrease in group achievement.
Second, as diversity increases, cooperation levels of agents at high and low risk respectively increase and decrease. However, while the changes in risk exposure are symmetrical, the changes in cooperation are not. The increase in cooperation of one class is always smaller than the accompanied decrease in cooperation of the other class, which raises significant target achievement difficulties.
Third, we showed that \acrshort{rl} populations converge to more egalitarian solutions among the two classes with respect to their class-based counterparts.
Finally, we discussed how learning in large \acrshort{rl}  populations may help in avoiding defective Nash equilibria.

We recall that risk diversity can emerge from a misalignment in either risk perception or risk exposure.
In the case of risk perception diversity, our results highlight the need to align risk perceptions among individuals --- using education for example \cite{lee2015predictors} --- to improve a population's ability in collectively reaching a target.
However, if diversity in risk exposure relates to geographic locations, health problems, or other non modifiable variables, collective success demands altruistic actions from agents who may not directly benefit from cooperating.
When agents are at very low risk, cooperative actions cannot be enforced using communication, retaliation or other classical solutions for cooperation in symmetrical social dilemmas.
Mixing individualistic and social qualities in agents is necessary to achieve cooperative AI under risk diversity.
For fully individualistic agents, allowing inter-agent contracts and bargains can be a way for selfish cooperation to emerge.
This requires the understanding of the payoffs of the game, the capacity to develop win-win proposals, and the ability to implement those contracts (i.e, the ability to receive and offer rewards or incentives from and to other agents).



\begin{ack}
This work was partially supported by FCT-Portugal (UIDB/50021/2020, PTDC/MAT-APL/6804/2020, and PTDC/CCI-INF/7366/2020). This work has also received funding from the European Union’s H2020 program (grant 76595).
\end{ack}



\bibliographystyle{plain} 
\bibliography{bibliography}


\appendix

\section{Appendix}

\subsection{Game definition - requirements for a social dilemma}
\label{appendix:game definition}

Social dilemmas rise from a misalignment of individual and collective interests generated by specific tensions in the payoff function \cite{macy2002learning}. More specifically, mutual cooperation should always be preferred over a unilateral cooperation and a mutual defection.
However, there should also always be either greed or fear that drives agents to defect to either exploit their peer or protect themselves from exploitation.
Recall that in our game, a disaster is faced with probability $r_i$ by agent $i$ if the group fails to achieve the target threshold. 
A total cooperation always results in target achievement while a total defection 
always results in failure of target achievement.
As such, mutual cooperation yields a payoff $x_C =\log(1-c)$, whereas mutual defection yields with probability $r_i$ a payoff $\bar{x}_D=\log(1-p)$ and with probability $1-r_i$, a payoff $x_D=0$. 
To satisfy the conditions for a social dilemma, $x_C>(1-r_i)x_D + r_i \bar{x}_D$ which implies that $r_i > \frac{\log(1-c)}{\log(1-p)}$.
Additionally, the threshold $\mathbf{t}$ needs to be lower bounded by $cb$, otherwise a unilateral cooperation would also avoid the disaster and hence be as good as a mutual cooperation.
Finally, to incentivize agents to defect, the threshold needs to be achievable with less than a total cooperation ($\mathbf{t}<Ncb$), otherwise agents would have no motivation to free-ride.


\subsection{Agent learning algorithm}
\label{appendix:agent definition}

We train asynchronously agents of a population learning with the update rule in equation \ref{eq:update rule}. A comparison between synchronous and asynchronous learning showed no significant differences in results for players learning to play the Ultimatum Game \cite{santos2016dynamics}. Similar conclusions were also reached with populations facing collective risks \cite{merhej2021cooperation}.
At every update-step $k$, a group of $N$ agents is selected randomly from the population of $Z$ agents. The agents in this group engage in the game described in section \ref{sec:game definition}. Every player $i$ in the group chooses randomly one of the available actions following probabilities $\mathbf{p}_{i,k}$ that are derived by normalizing the propensity vector $\mathbf{q}_{i,k}$. The selected actions determine whether or not the target is achieved. If this is the case, then all agents avoid a disaster. 
Otherwise, the occurrence or not of a disaster for agent $i$, is sampled according to its risk exposure level $r_i$. The payoffs for each agent are then distributed according to Section~\ref{sec:game definition} after which all agents in the group update their propensity vectors. This is repeated for a total of $K$ update-steps. While training, we keep track of the number of times every agent in the population has been selected in a vector $\mathbf{u}$. Since the algorithm does not guarantee that all agents are chosen equally as many times, we define $K^{\prime}$, the minimum number of update-steps every agent needs to have performed before training is done. If after $K$ total update-steps, some agent still hasn't performed at least $K^{\prime}$ updates, then training continues until this condition is satisfied.

Because the payoffs added to the propensity vector $\mathbf{q}_{i,k}$ are negative, we choose the Softmax function to derive $p_{i,k}(A)$, the normalized propensities for each action. We have
\begin{equation}
\label{eq:propensity normalisation}
    p_{i,k}(A) = \frac{\exp(q_{i,k}(A))}{\sum_{A'\in\{C,D\}}\exp(q_{i,k}(A'))}.
\end{equation}

We initialize the propensity vectors $\mathbf{q}_{i,0}$ by sampling for each action, a random propensity value from a normal distribution $\mathcal{N}(\mu=0,\sigma=1)$.

During learning, we set the total number of update-steps to $K=2,500,000$ and impose a minimum number of $K^{\prime}=30,000$ updates for every agent. The forgetting parameter is set to $\phi=0.001$. All simulations are repeated for 5 runs.


\subsection{Static study}
\label{appendix:sec:risk ineq:static study}

\subsubsection{Class-based Nash}
\label{appendix:subsec:risk ineq:class-based Nash}

We follow an analogous reasoning to the one for finding class-based Nash strategies under wealth inequality in collective risks \cite{merhej2021cooperation}. 
We repeat the analysis while modifying what is necessary to accommodate homogeneous initial wealth and risk diversity.
We detail the steps of this procedure below:

\begin{figure*}
\centering
  \subfloat[$r=0.1$]{
	\begin{minipage}[c][1\width]{
	   0.45\textwidth}
	   \centering
	   \includegraphics[width=1\textwidth]{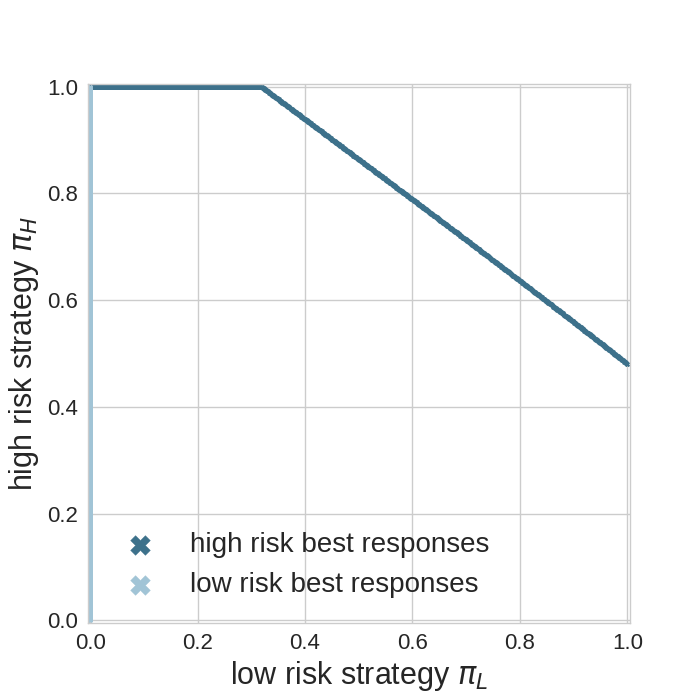}
	\end{minipage}}
 \\
  \subfloat[$r=0.3$]{
	\begin{minipage}[c][1\width]{
	   0.45\textwidth}
	   \centering
	   \includegraphics[width=1\textwidth]{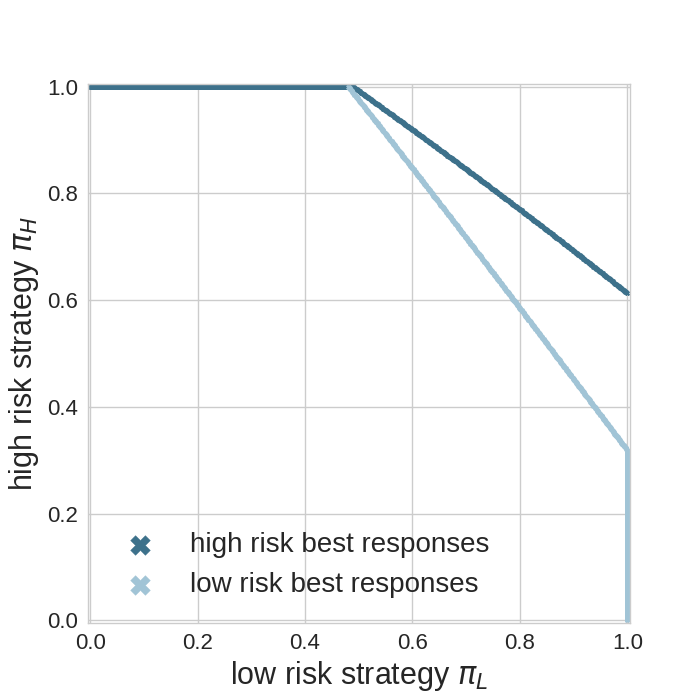}
	\end{minipage}}
 \hfill	
  \subfloat[$r=0.5$]{
	\begin{minipage}[c][1\width]{
	   0.45\textwidth}
	   \centering
	   \includegraphics[width=1\textwidth]{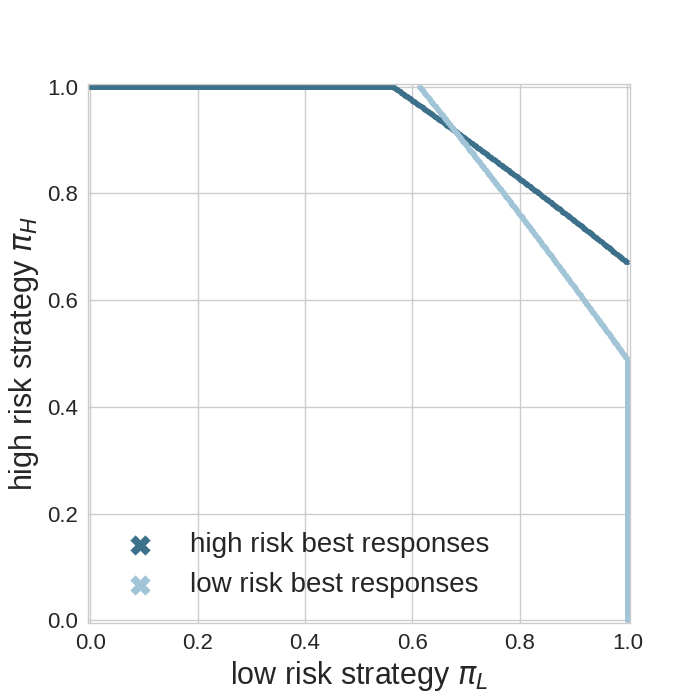}
	\end{minipage}}
	\hfill 	
  \subfloat[$r=0.7$]{
	\begin{minipage}[c][1\width]{
	   0.45\textwidth}
	   \centering
	   \includegraphics[width=1\textwidth]{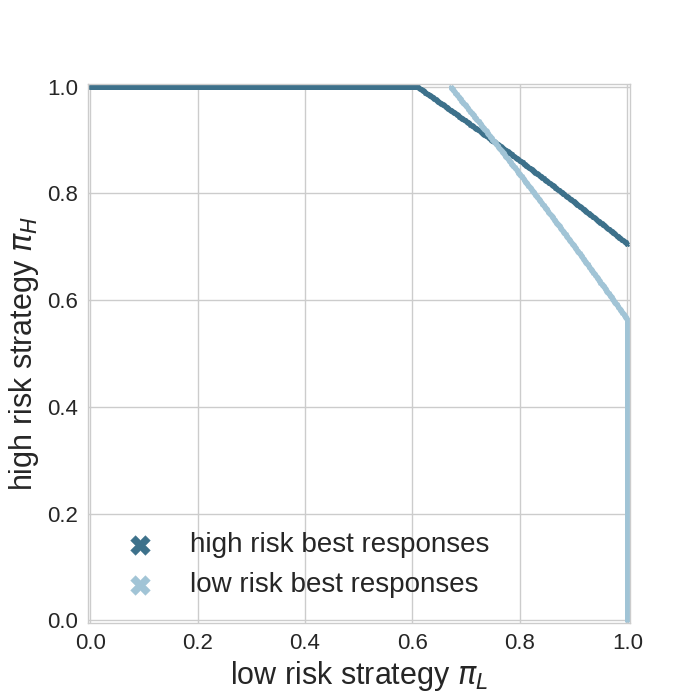}
	\end{minipage}}
\hfill	
   \subfloat[$r=0.9$]{
 	\begin{minipage}[c][1\width]{
 	   0.45\textwidth}
 	   \centering
 	   \includegraphics[width=1\textwidth]{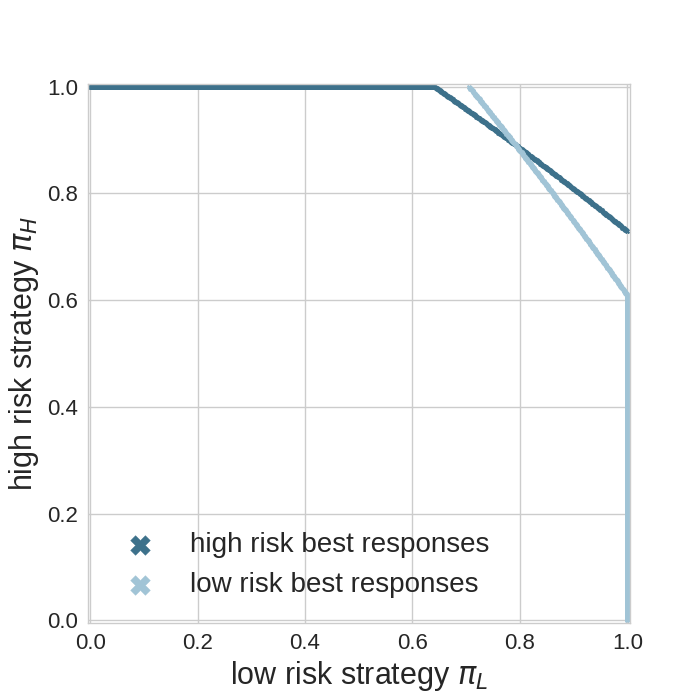}
 	\end{minipage}}
\caption{Plots used in extracting class-based Nash equilibrium points for games with a risk diversity $\delta=0.1$ and different average risk values.}
\label{fig:nash risk ineq: delta=0.1}
\end{figure*}

Consider a group of $N-1$ individuals and denote by $n_L$ and $n_H$ respectively, the number of agents at low and high risk within this group where $n_L \in \{ 0,1,...,N-1\}$ and $n_H = N -1-n_L$. Let $n_L^c$ be the number of players at low risk that actually contribute to the pool i.e. $n_L^c \in \{0,1,...,n_L\}$ and $n_H^c$ be the number of contributors at high risk in the group i.e. $n_H^c \in \{0,1,...,n_H\}$. Hence, a total number of $(n_L+1)\times(n_H+1)$ different combinations of group contributions are possible.

The probability $P^{n_L}(n_L^c,n_H^c)$ that each of these possible configurations occur in a group of $n_L$ agents at low risk follows a binomial law and depends on $\pi_L$ and $\pi_H$. 
\begin{equation}
\begin{aligned}
P^{n_L}(n_L^c,n_H^c) =& \binom{n_L}{n_L^c}\pi_L^{n_L^c} (1-\pi_L)^{n_L-n_L^c} \binom{n_H}{n_H^c} \pi_H^{n_H^c} (1-\pi_H)^{n_H-n_H^c}
\end{aligned}
\end{equation}

Let $i$ be the $N^{th}$ player to join the group.
Player $i$ will now choose to contribute with probability $\pi_L$ if he's at low risk or with probability $\pi_H$ if he is at high risk. Denote by $A^D$ the action of defecting and $A^C$ the action of contributing.
Denote by $\mathcal{S}_{A^D}$  the set of configurations that achieve the threshold without the need of $i$'s contribution. Mathematically, $\mathcal{S}_{A^D}=\{\forall \ (n_L^c,n_H^c) \in \{0,1,...,n_L\}\times\{0,1,...,n_H\} |n_L^cbc+n_H^cbc \geq Mbc\}$. Identically, denote by $\mathcal{S}_{A^C}$
the set of configurations that can achieve the threshold if $i$ contributes.
The probability of the group achieving the threshold given that $i$ chose action $a \in \{A^D, A^C\}$ is given by the sum of the probabilities of the events in $\mathcal{S}_{A^D}$ and $\mathcal{S}_{A^C}$ respectively.
 \begin{equation}
     \begin{aligned}
     P^{n_L}(\mathbf{t}|a) = \sum_{(n_L^c,n_H^c)\in \mathcal{S}_a} P^{n_L}(n_L^c,n_H^c)
     \end{aligned}
 \end{equation}

Since the game is probabilistic, the probability of a player $i$ avoiding a disaster given that he chose action $a$ is given by equations \ref{eq:p_success low risk} for players at low risk and equations \ref{eq:p_success high risk} for players at high risk.
\begin{equation}
\label{eq:p_success low risk}
\begin{aligned}
    P_L^{n_L}(\text{success}|a) &= P^{n_L}(\mathbf{t}|a) + (1-r_L)P^{n_L}(\neg \mathbf{t}|a) \\
    P_L^{n_L}(\text{failure}|a)&=1-P_L^{n_L}(\text{success}|a)
\end{aligned}
\end{equation}

\begin{equation}
\label{eq:p_success high risk}
\begin{aligned}
    P_H^{n_L}(\text{success}|a) &= P^{n_L}(\mathbf{t}|a) + (1-r_H)P^{n_L}(\neg \mathbf{t}|a) \\
    P_H^{n_L}(\text{failure}|a)&=1-P_H^{n_L}(\text{success}|a)
\end{aligned}
\end{equation}

We can now write the expected payoff functions of player $i$ depending on whether he's at low or high risk. Let $\mathcal{H}_L^{n_L}(\pi_L, \pi_H)$ and $\mathcal{H}_H^{n_L}(\pi_L, \pi_H)$ be the respective expected payoff functions of agents at low and high risk involved in a game with $n_L$ players at low at risk and where all agents at low risk follow strategy $\pi_L$ and all agents at high risk follow strategy $\pi_H$. The expected payoff of an agent depends on whether the game was successful or not and whether he contributed or not to the common pool. We have
\begin{equation}
\label{eq:payoff low risk}
\begin{aligned}
\mathcal{H}^{n_L}_L(\pi_L, \pi_H) =&\pi_L [P_L^{n_L}(\text{success} | A^C)x_C + P_L^{n_L}(\text{failure} | A^C)\bar{x}_C ] +\\
&(1-\pi_L)[P_L^{n_L}(\text{success} | A^D)x_D + P_L^{n_L}(\text{failure} | A^D)\bar{x}_D]
\end{aligned}
\end{equation}
\begin{equation}
\label{eq:payoff high risk}
\begin{aligned}
\mathcal{H}^{n_L}_H(\pi_L, \pi_H) =&\pi_H[P_H^{n_L}(\text {success} | A^C)x_C + P_H^{n_L}(\text{failure} | A^C)\bar{x}_C]+ \\
&(1-\pi_H)[P_H^{n_L}(\text{success} | A^D)x_D + P_H^{n_L}(\text{failure} | A^D)\bar{x}_D]
\end{aligned}
\end{equation}
where $x_C$, $\bar{x}_C$, $x_D$ and $\bar{x}_D$ are the payoffs described in Section~\ref{sec:game definition}.

Finally, as groups are sampled randomly, the expected payoff accounting for the probability of an agent to find himself in a group with $n_L$ agents at low risk is

\begin{equation}
\label{eq:average return risk ineq}
    \begin{aligned}
    \mathcal{H}_L(\pi_L, \pi_H) &= \sum_{n_L=0}^ {N-1}\frac{\binom{Z_L-1}{n_L}\binom{Z-Z_L}{N-n_L-1}}{\binom{Z-1}{N-1}} \mathcal{H}^{n_L}_L(\pi_L, \pi_H)\\
    \mathcal{H}_H(\pi_L, \pi_H) &= \sum_{n_L=0}^ {N-1}\frac{\binom{Z_L}{n_L}\binom{Z-Z_L-1}{N-n_L-1}}{\binom{Z-1}{N-1}} \mathcal{H}^{n_L}_H(\pi_L, \pi_H).
    \end{aligned}
\end{equation}

Both players at low and high risk exposure aim at maximizing their respective payoff functions $\mathcal{H}_L$ and $\mathcal{H}_H$. A Nash equilibrium $(\pi_L^*, \pi_H^*)$ satisfies
\begin{equation}
\label{eq:Nash risk ineq}
    \begin{aligned}
    \mathcal{H}_L(\pi_L^*, \pi_H^*) &\geq \mathcal{H}_L(\pi_L, \pi_H^*) \quad &\forall \ \pi_L\in [0,1]\\
    \mathcal{H}_H(\pi_L^*, \pi_H^*) &\geq \mathcal{H}_H(\pi_L^*, \pi_H)\quad &\forall \ \pi_H \in [0,1]
    \end{aligned}
\end{equation}

\begin{figure*}[h!]
\centering
  \subfloat[$\delta=0.2$]{
	\begin{minipage}[c][1\width]{
	   0.45\textwidth}
	   \centering
	   \includegraphics[width=1\textwidth]{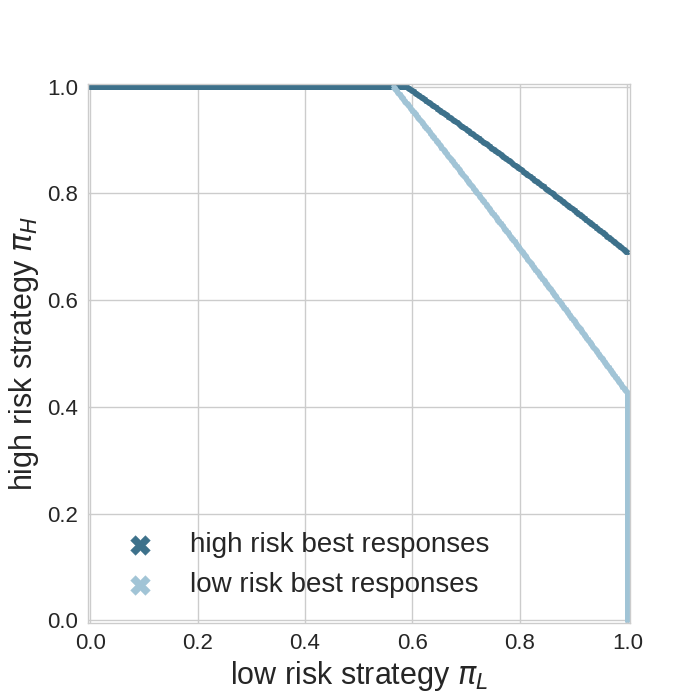}
	\end{minipage}}
 \hfill	
  \subfloat[$\delta=0.3$]{
	\begin{minipage}[c][1\width]{
	   0.45\textwidth}
	   \centering
	   \includegraphics[width=1\textwidth]{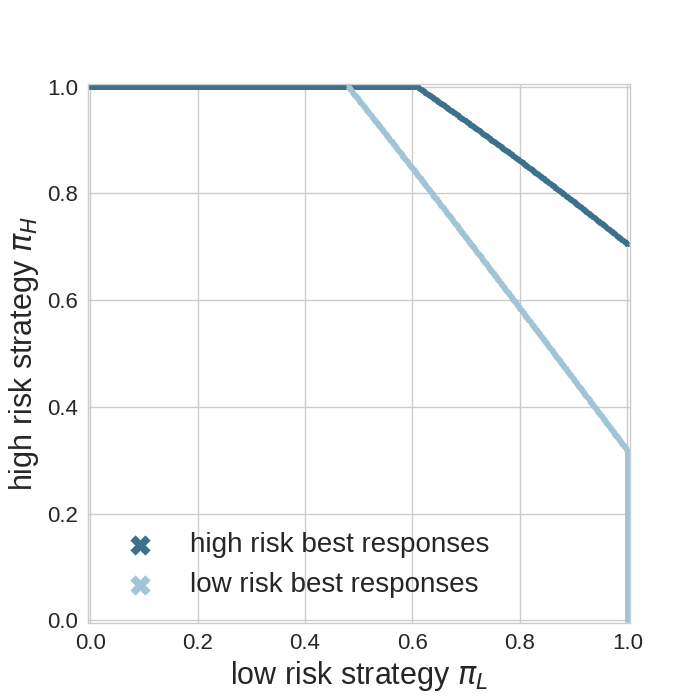}
	\end{minipage}}
	\hfill 	
  \subfloat[$\delta=0.4$]{
	\begin{minipage}[c][1\width]{
	   0.45\textwidth}
	   \centering
	   \includegraphics[width=1\textwidth]{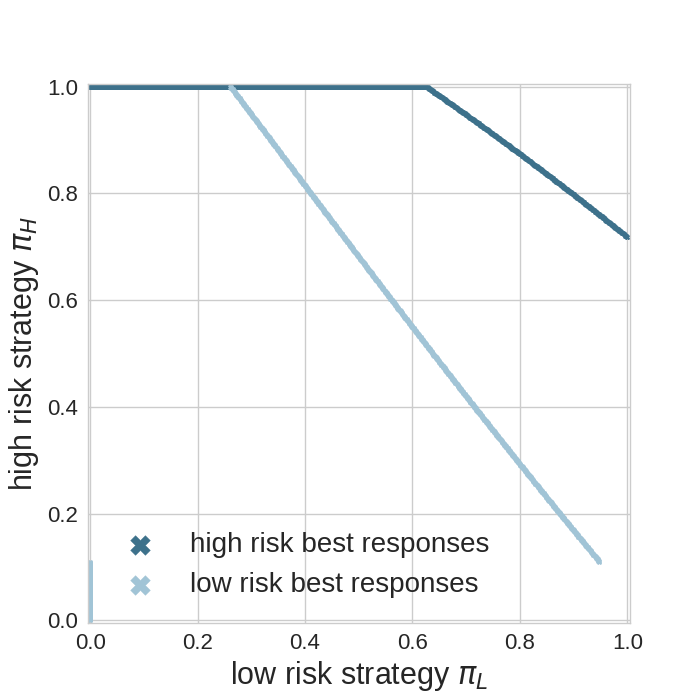}
	\end{minipage}}
\hfill	
   \subfloat[$\delta=0.5$]{
 	\begin{minipage}[c][1\width]{
 	   0.45\textwidth}
 	   \centering
 	   \includegraphics[width=1\textwidth]{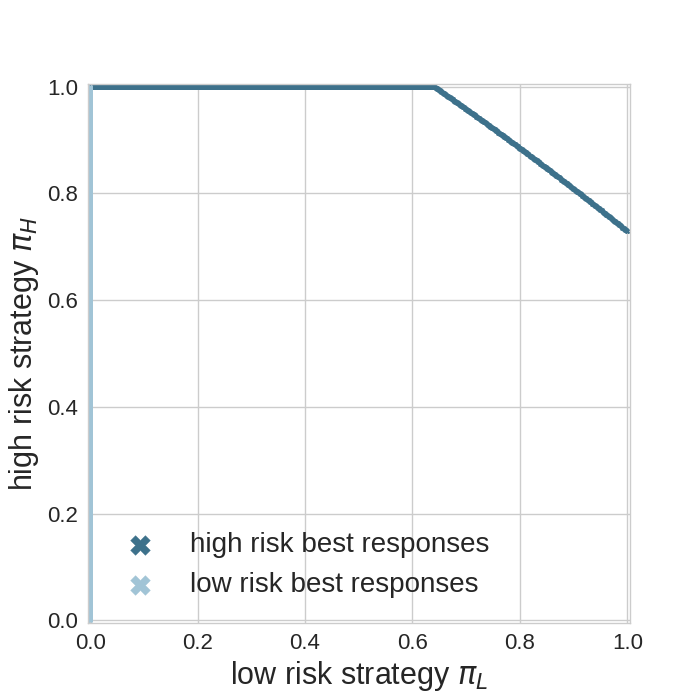}
 	\end{minipage}}
\caption{Plots used in extracting class-based Nash equilibrium points for games with an average risk value $r=0.5$ and different risk diversity values $\delta$}
\label{fig:nash risk ineq: r=0.5}
\end{figure*}

Again, we rely on a graphical method and discretize the domain of $\pi_L$ and $\pi_H$ into intervals of length $\epsilon =0.001$. We calculate the corresponding payoff $\mathcal{H}_L$ and $\mathcal{H}_H$ over the space of possible $(\pi_L,\pi_H)$.
Referring to equations \ref{eq:average return risk ineq}, we plot for every $\pi_H$, $L$'s best response $\pi^{BR}_L$ i.e. $\pi^{BR}_L \ s.t. \ \mathcal{H}_L(\pi^{BR}_L, \pi_H)$ is maximized and similarly for every $\pi_L$, $H$'s optimal response $\pi^{BR}_H $.
The intersections of the hence formed lines represent class-based Nash equilibrium points. We extract these points for different game configurations in each of our two scenarios: on one hand, for different average risk values $r$ with a fixed risk diversity $\delta=0.1$ (Figure~\ref{fig:nash risk ineq: delta=0.1}), and on the other hand, for different risk diversity factors $\delta$ and a fixed average risk factor $r=0.5$ (Figure~\ref{fig:nash risk ineq: r=0.5}).

\subsubsection{Class-based maximum welfare}
\label{appendix:subseb:risk ineq:max welfare}

\begin{figure*}[h!]
\centering
  \subfloat[$r=0.1$]{
	\begin{minipage}[c][0.8\width]{
	   0.45\textwidth}
	   \centering
	   \includegraphics[width=1\textwidth]{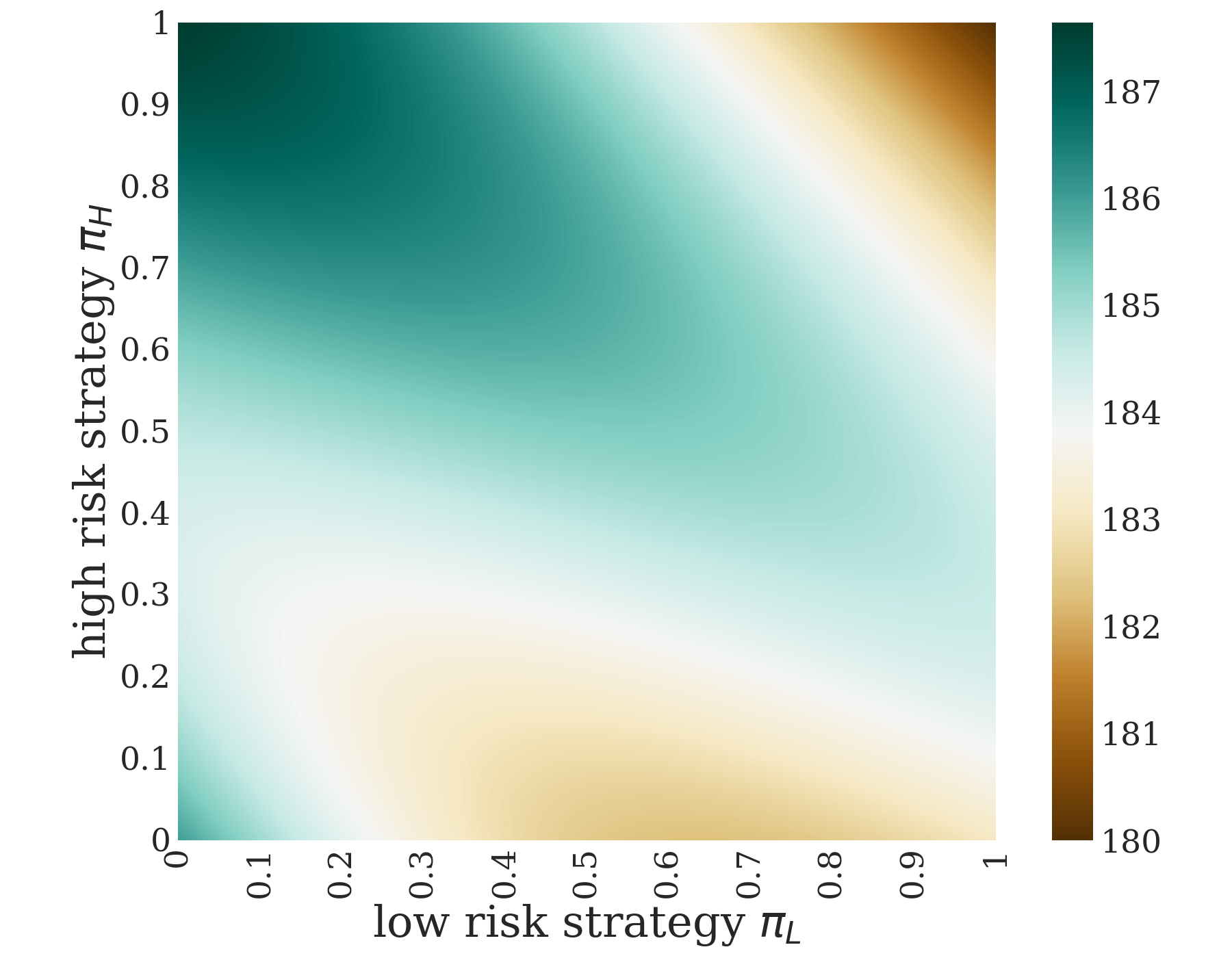}
	\end{minipage}}
 \hfill \\	
  \subfloat[$r=0.3$]{
	\begin{minipage}[c][0.8\width]{
	   0.45\textwidth}
	   \centering
	   \includegraphics[width=1\textwidth]{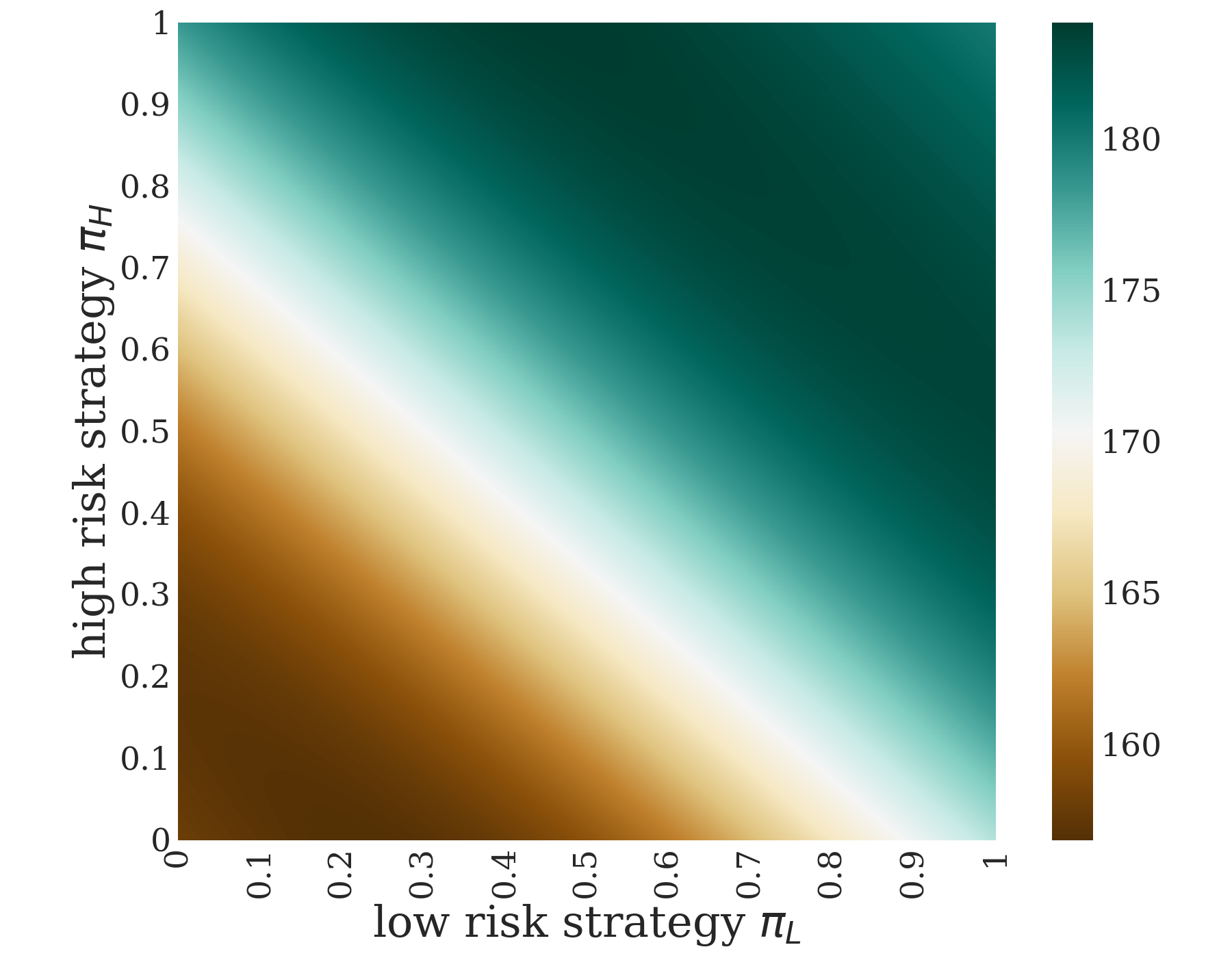}
	\end{minipage}}
 \hfill	
  \subfloat[$r=0.5$]{
	\begin{minipage}[c][0.8\width]{
	   0.45\textwidth}
	   \centering
	   \includegraphics[width=1\textwidth]{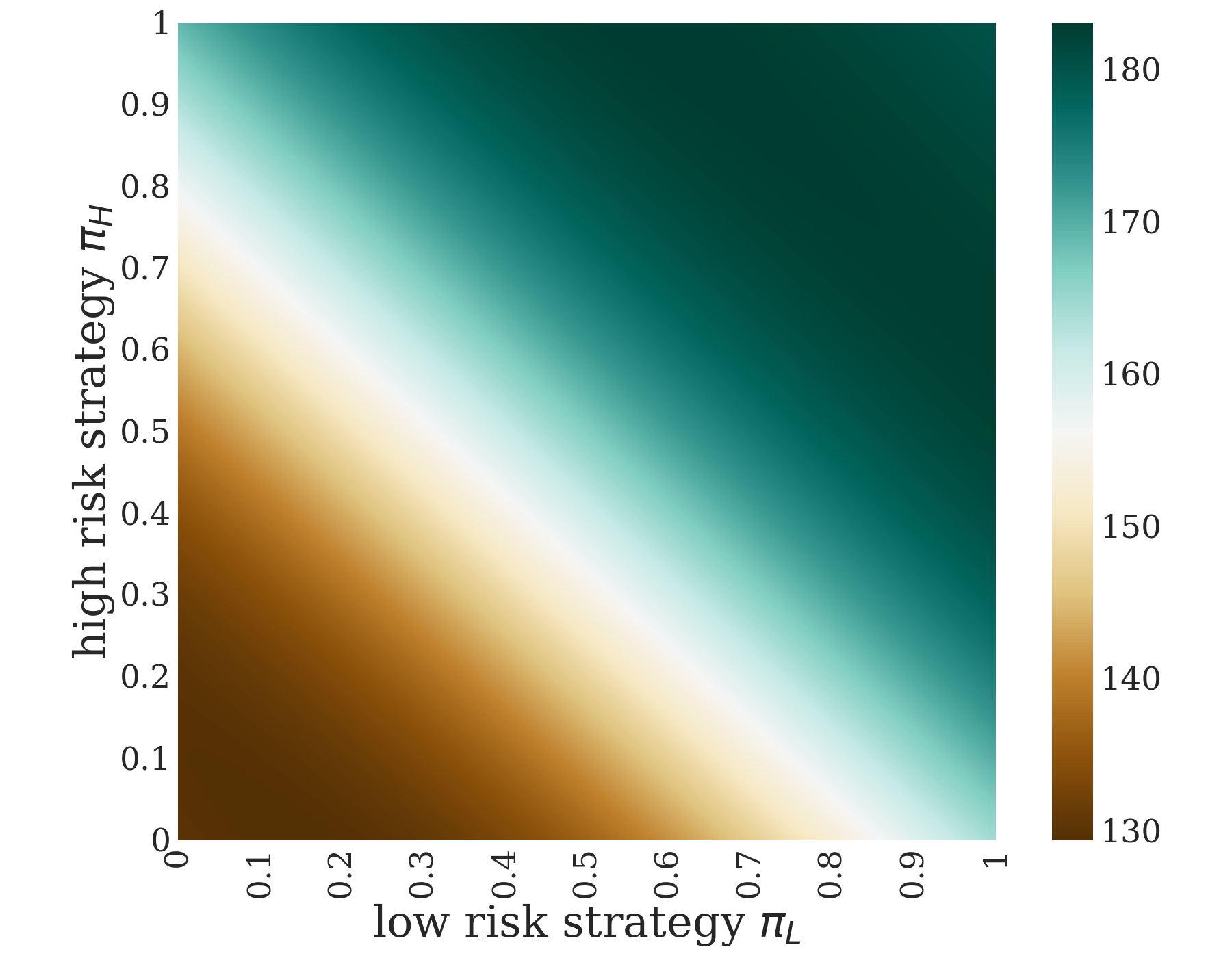}
	\end{minipage}}
	\hfill 	
  \subfloat[$r=0.7$]{
	\begin{minipage}[c][0.8\width]{
	   0.45\textwidth}
	   \centering
	   \includegraphics[width=1\textwidth]{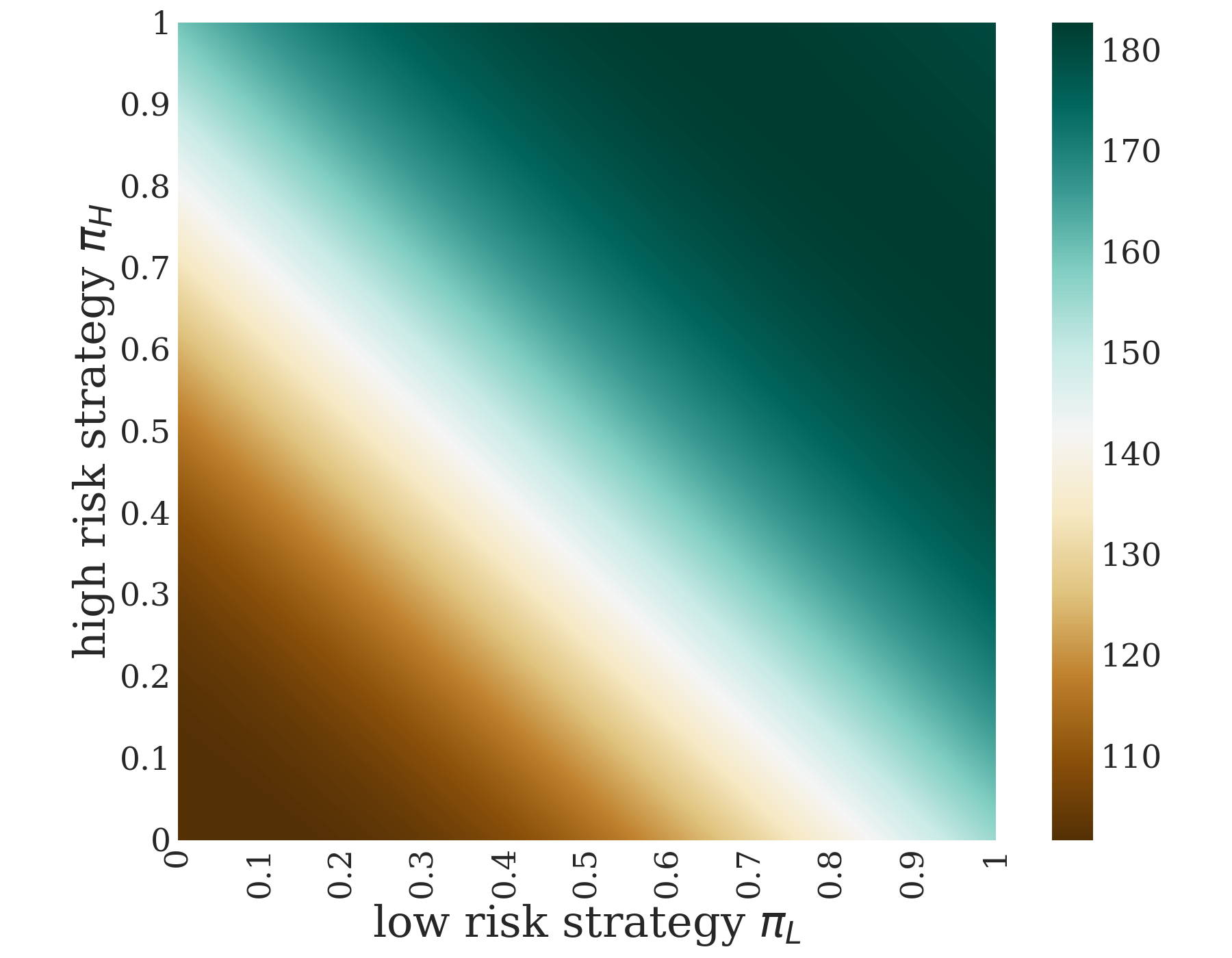}
	\end{minipage}}
 \hfill	
   \subfloat[$r=0.9$]{
 	\begin{minipage}[c][0.8\width]{
 	   0.45\textwidth}
 	   \centering
 	   \includegraphics[width=1\textwidth]{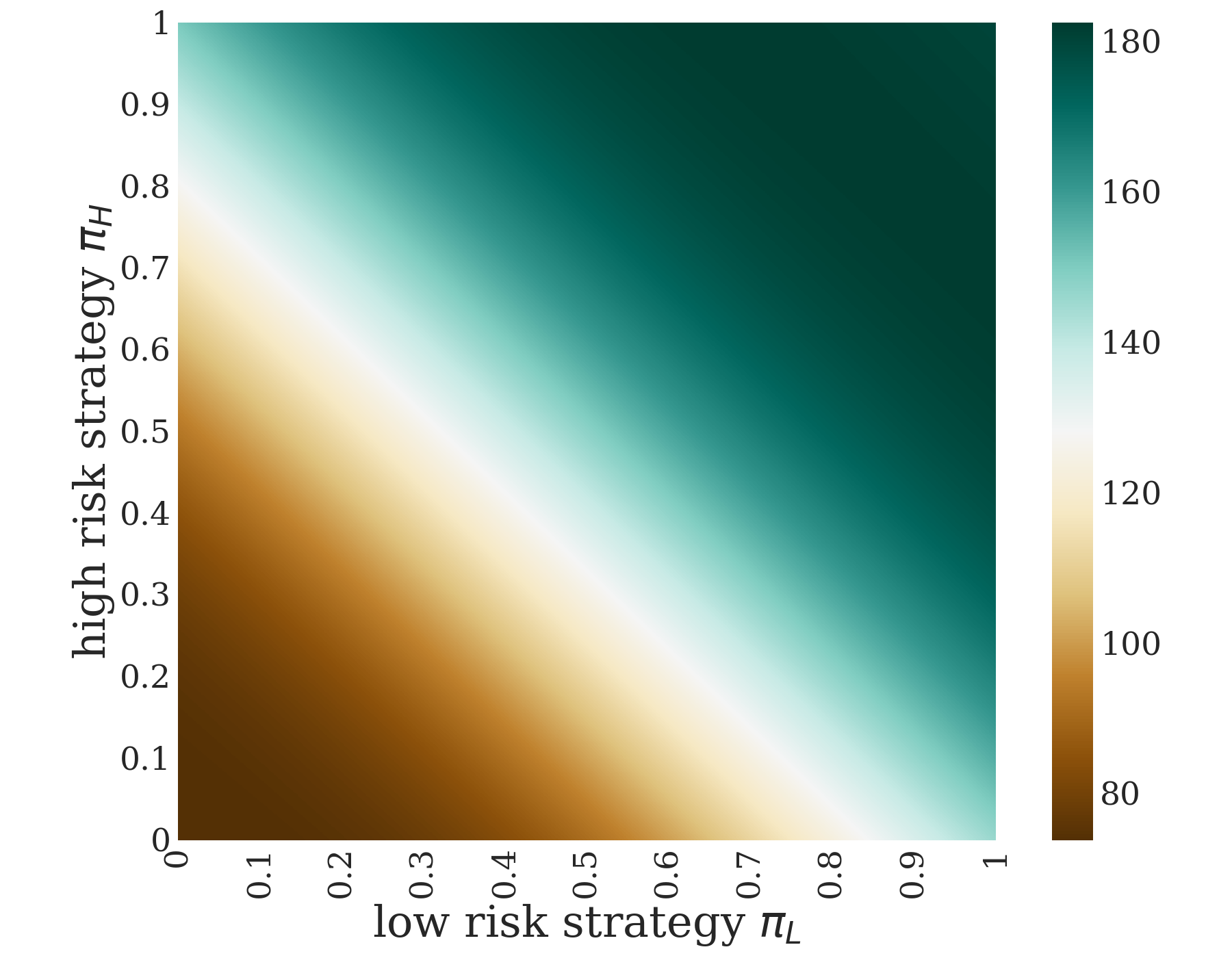}
\end{minipage}}
\caption{Heat-maps of total population welfare as a function of the chosen strategies by individuals at high and low risk of facing a disaster. In all plots, $\delta=0.1$.}
\label{fig:max welfare risk ineq:d=0.1}
\end{figure*}

\begin{figure*}[h!]
  \subfloat[$\delta=0.2$]{
	\begin{minipage}[c][0.8\width]{
	   0.45\textwidth}
	   \centering
	   \includegraphics[width=1\textwidth]{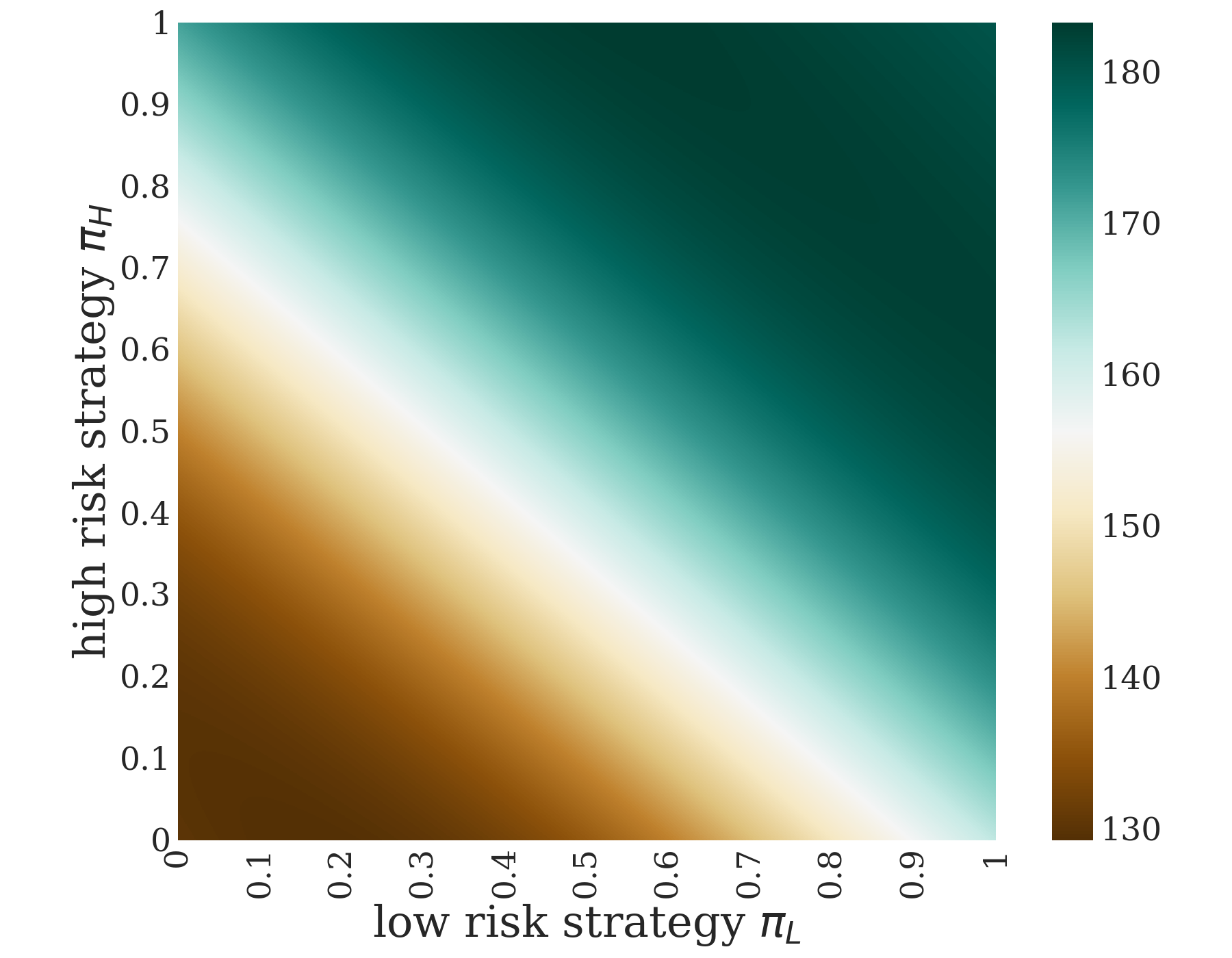}
	\end{minipage}}
 \hfill	
  \subfloat[$\delta=0.3$]{
	\begin{minipage}[c][0.8\width]{
	   0.45\textwidth}
	   \centering
	   \includegraphics[width=1\textwidth]{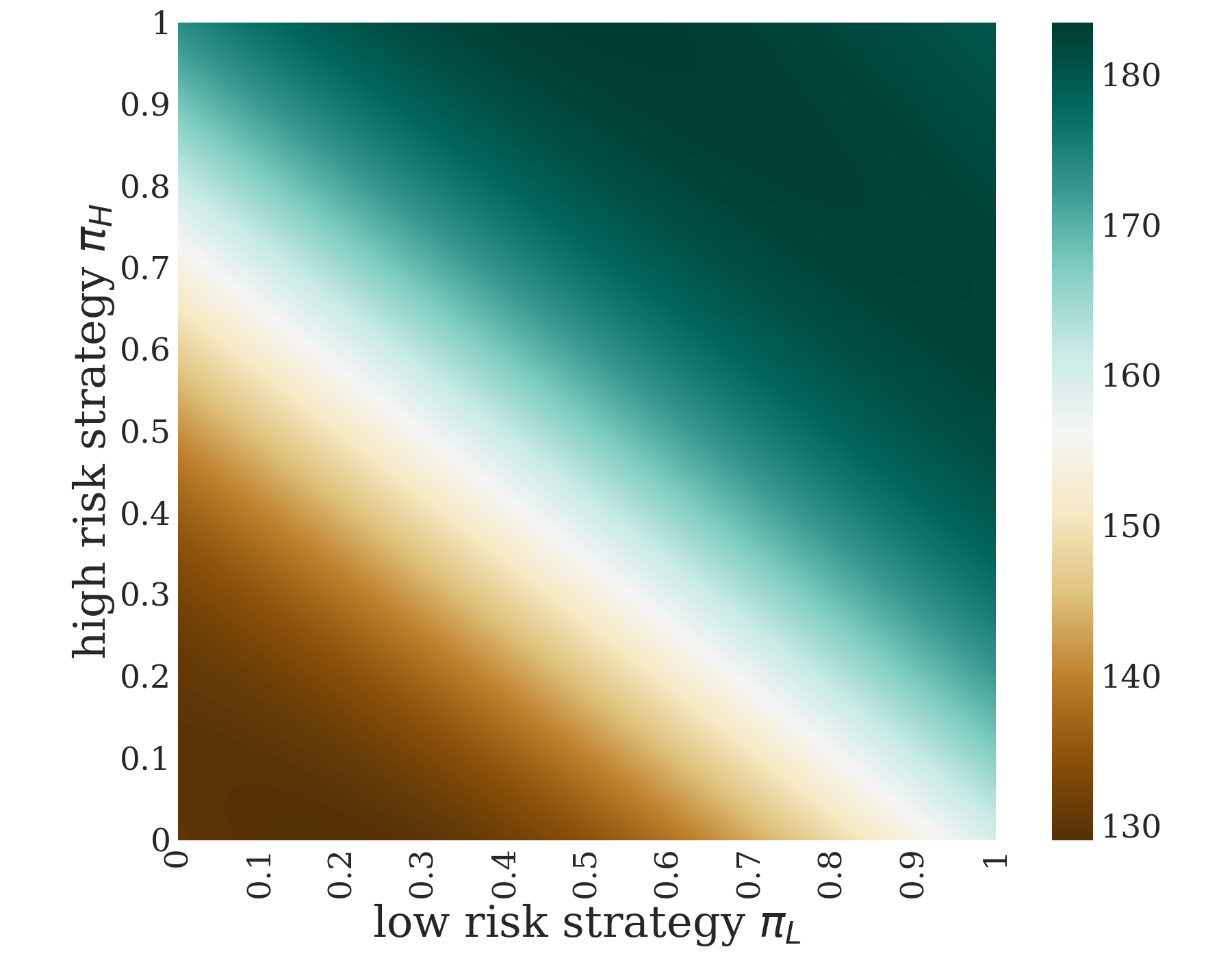}
	\end{minipage}}
	\hfill 	
  \subfloat[$\delta=0.4$]{
	\begin{minipage}[c][0.8\width]{
	   0.45\textwidth}
	   \centering
	   \includegraphics[width=1\textwidth]{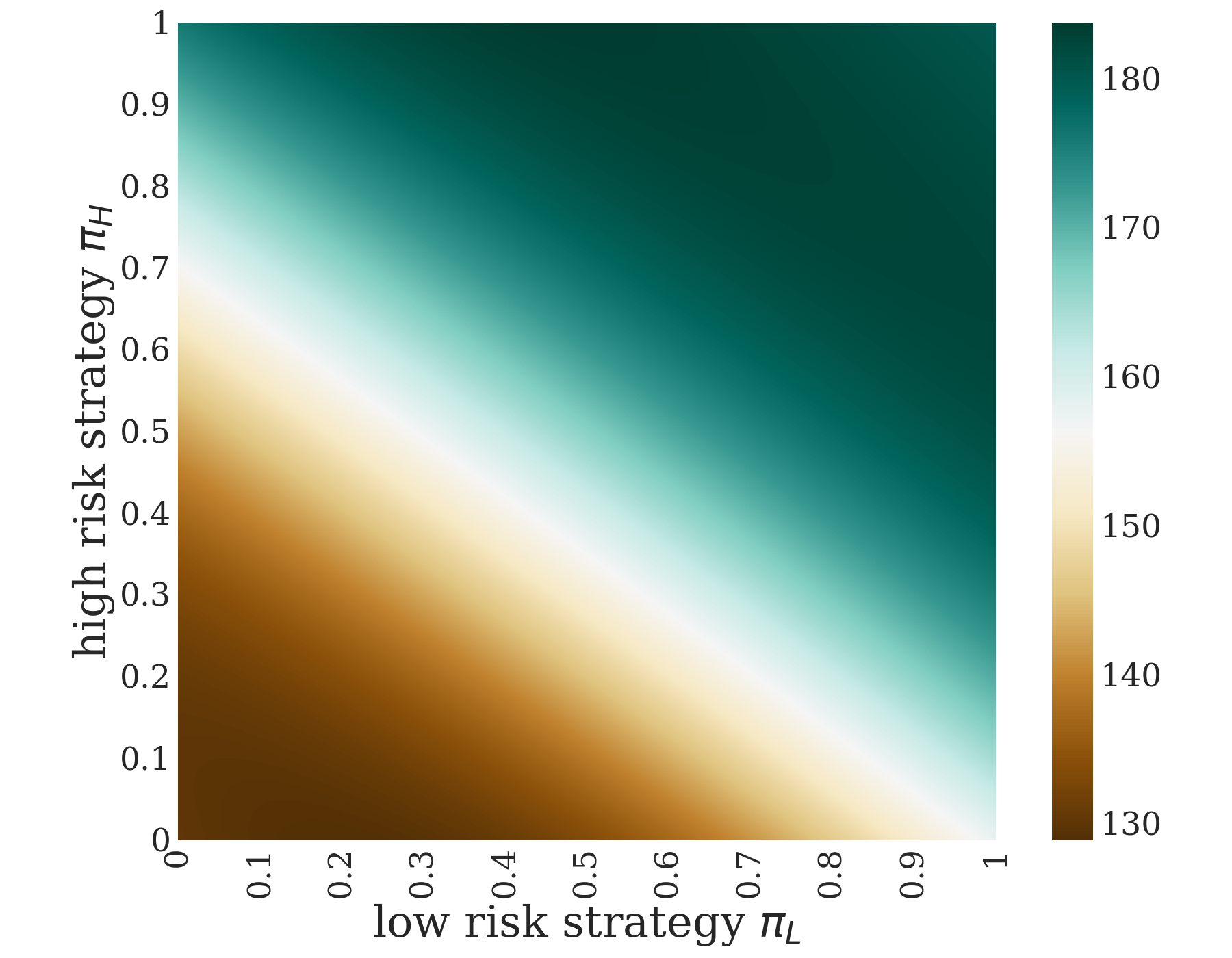}
	\end{minipage}}
 \hfill	
   \subfloat[$\delta=0.5$]{
 	\begin{minipage}[c][0.8\width]{
 	   0.45\textwidth}
 	   \centering
 	   \includegraphics[width=1\textwidth]{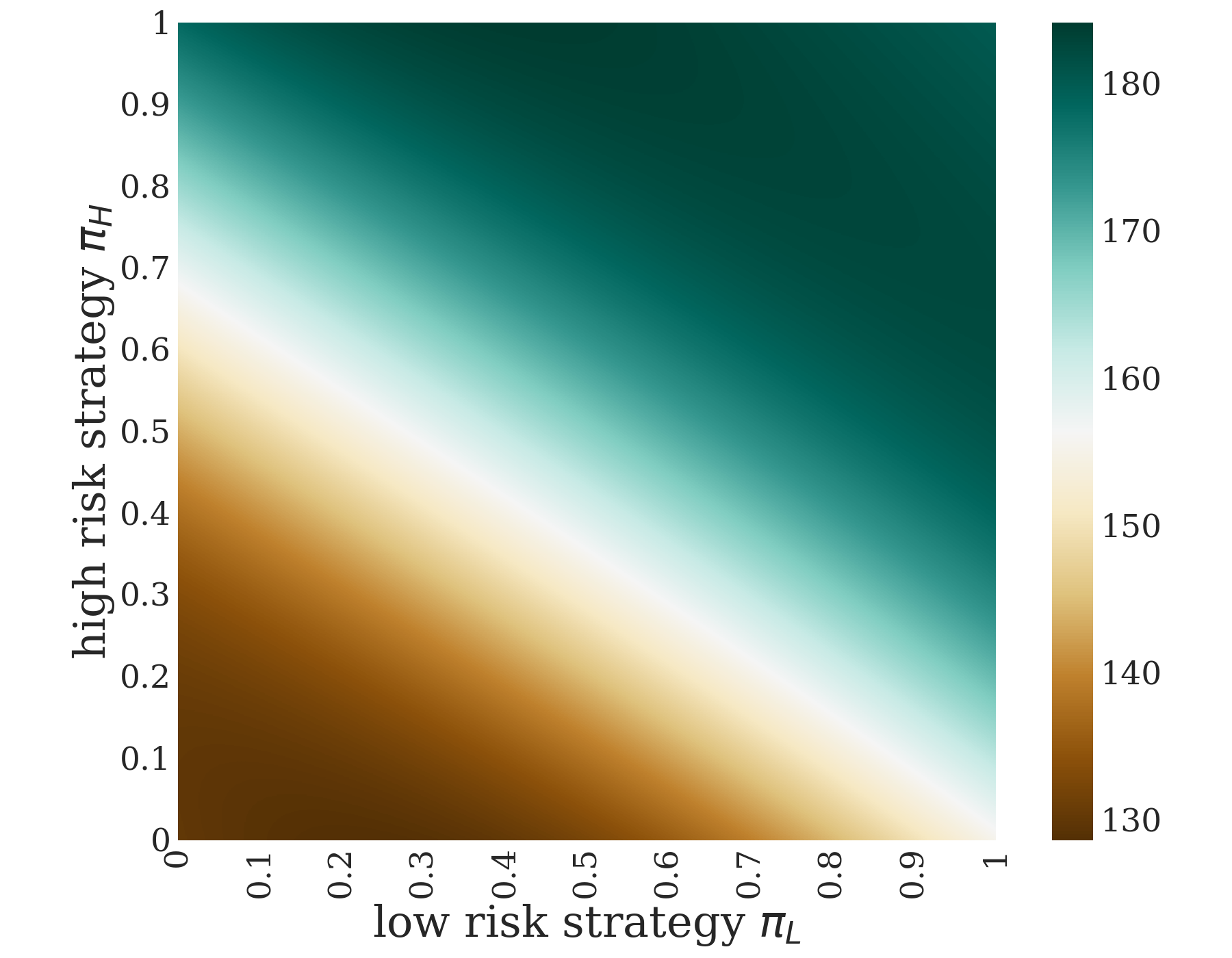}
\end{minipage}}
\caption{Heat-maps of total population welfare as a function of the chosen strategies by individuals at high and low risk. In all plots, the average risk of the population $r=0.5$.}
\label{fig:max welfare risk ineq:r=0.5}
\end{figure*}

When evaluating the expected return for the population, we do not look at the relative cost a loss has on an individual, but rather at the absolute impact it has on the population.
We modify the value of the log-utility returns $x_C$, $\bar{x}_C$, $x_D$ and $\bar{x}_D$ in Equations \ref{eq:payoff low risk} and \ref{eq:payoff high risk} and replace them by a linear utility.
A successful cooperation from an agent costs the society $x_C=-cb$ and a failed cooperation costs $\bar{x}_C=-cb-(1-c)pb$.
Similarly, a successful defection costs nothing $x_D=0$ whereas a failed defection incurs a cost of $\bar{x}_D=-pb$.
Then, using Equations \ref{eq:average return risk ineq}, we build a heat-map with the average population wealth for every combination of $\pi_L$ and $\pi_H$ strategies. 
Figure \ref{fig:max welfare risk ineq:d=0.1} illustrates some of the heat-maps obtained for different risk values and $\delta=0.1$, while Figure~\ref{fig:max welfare risk ineq:r=0.5} illustrates heat-maps obtained for different $\delta$ values and an average population risk factor $r=0.5$.
Dark green colors represent solutions maximizing social welfare.
We observe that the higher the risk factor, the lower the maximum social welfare obtained (see color bars).
In all cases, individuals at high risk are recommended to cooperate more than those at low risk.

\end{document}